# Fourier - transform spectroscopy of $^{13}C^{17}O$ and deperturbation analysis of the $A^1\Pi$ ($\upsilon$ = 0 - 3) levels

R. Hakalla[1,*], M. L. Niu[2], R. W. Field[3], A. N. Heays[2], E. J. Salumbides[2,4], G. Stark[5], J. R. Lyons[6], M. Eidelsberg[7], J. L. Lemaire[8], S. R. Federman[9], N. de Oliveira[10], and W. Ubachs[2]

[1] *Materials Spectroscopy Laboratory, Department of Experimental Physics, Faculty of Mathematics and Natural Science, University of Rzeszów, ul. Prof. S. Pigonia 1, 35-959 Rzeszów, Poland.*
[2] *Department of Physics and Astronomy, and LaserLaB, Vrije Universiteit, De Boelelaan 1081, 1081 HV Amsterdam, Netherlands.*
[3] *Department of Chemistry, Massachusetts Institute of Technology, Cambridge, MA 02139, USA.*
[4] *Department of Physics, University of San Carlos, Cebu City 6000, Philippines.*
[5] *Department of Physics, Wellesley College, Wellesley, MA 02481, USA.*
[6] *School of Earth and Space Exploration, Arizona State University, PO Box 871404, Tempe, AZ 85287, USA.*
[7] *Observatoire de Paris, LERMA, UMR 8112 du CNRS, 92195 Meudon, France.*
[8] *Institut des Sciences Moléculaires d'Orsay (ISMO), CNRS - Université Paris-Sud, UMR 8214, 1405 Orsay, France (previously at Paris Observatory, LERMA).*
[9] *Department of Physics and Astronomy, University of Toledo, Toledo, OH 43606, USA.*
[10] *Synchrotron SOLEIL, Orme de Merisiers, St. Aubin, BP 48, F-91192 Gif sur Yvette Cedex, France.*

A B S T R A C T

The high-resolution $B^1\Sigma^+ \rightarrow A^1\Pi$ (0, 0) and (0, 3) emission bands of the less-abundant $^{13}C^{17}O$ isotopologue have been investigated by Fourier-transform spectroscopy in the visible region using a Bruker IFS 125HR spectrometer at an accuracy 0.003 cm$^{-1}$. These spectra are combined with high-resolution photoabsorption measurements of the $^{13}C^{17}O$ $B^1\Sigma^+ \leftarrow X^1\Sigma^+$ (0, 0), $B^1\Sigma^+ \leftarrow X^1\Sigma^+$ (1, 0) and $C^1\Sigma^+ \leftarrow X^1\Sigma^+$ (0, 0) bands recorded with an accuracy of 0.01 cm$^{-1}$ using the vacuum ultraviolet Fourier-transform spectrometer, installed on the DESIRS beamline at the SOLEIL synchrotron. In the studied 17,950 - 22,500 cm$^{-1}$ and 86,800 - 92,100 cm$^{-1}$ regions, 480 transitions have been measured. These new experimental data were combined with data from the C → A and B → A systems, previously analyzed in $^{13}C^{17}O$. The frequencies of 1003 transitions derived from 12 bands were used to analyze the perturbations between the $A^1\Pi$ ($\upsilon$ = 0 - 3) levels and rovibrational levels of the $d^3\Delta_i$, $e^3\Sigma^-$, $a'^3\Sigma^+$, $I^1\Sigma^-$ and $D^1\Delta$ states as well as to a preliminary investigation of weak irregularities that appear in the $B^1\Sigma^+$ ($\upsilon$ = 0) level. Deperturbed molecular constants and term values of the $A^1\Pi$ state were obtained. The spin-orbit and *L*-uncoupling interaction parameters as well as isotopologue-independent spin-orbit and rotation-electronic perturbation parameters were derived.

*corresponding author: rhakalla@ur.edu.pl

## 1. Introduction

The spectroscopy of the carbon monoxide (CO) molecule remains of central interest to a variety of subfields in science, from the submm [1,2] to the X-ray ranges [3]. It is of relevance, inter alia, to many astronomical environments of the interstellar medium (ISM) [4-9], planetary [10] and exoplanetary atmospheres [11], cosmology [12], as well as in combustion processes [13], atmospheric physics [14], and plasma science [15,16]. Various CO isotopologues are often used as tracers to determine the content of molecular clouds and to map the distribution of matter [17-19]. An important description of modeling CO photochemistry was made by Visser et al. [20]. The large CO abundance gives a detectable signal for even the least-abundant isotopologues [21-27], including $^{13}C^{17}O$ [23,27]. The first observation of $^{13}C^{17}O$ in the ISM was made in 2001 by Bensch et al. [23] in the $\rho$-Ophiuchi molecular cloud with the ESO Submillimetre Telescope (SEST) at La Silla, Chile. Under laboratory conditions the $^{13}C^{17}O$ isotopologue was studied several times [14,28-31], recently observed in the vacuum ultraviolet (VUV) by Lemaire et al. [32], and investigated in the visible (VIS) by Hakalla et al. [33,34] and Hakalla [35].

The $A^1\Pi$ state of carbon monoxide is among the most extensively studied electronic states in molecular spectroscopy. While it is important in a variety of applications, the $A^1\Pi$ state is a benchmark for the study of multistate perturbations [36]. These perturbations are caused by rovibrational levels of the $e^3\Sigma^-$, $a'^3\Sigma^+$, and $d^3\Delta_i$ triplet states as well as the $I^1\Sigma^-$ and $D^1\Delta$ singlet states. A systematic classification of the perturbations occurring in the $A^1\Pi$ state of the main $^{12}C^{16}O$ isotopologue was conducted by Krupenie [37] and Simmons et al. [38]. A considerable contribution to the identification and classification of the $A^1\Pi$ state perturbations has been made by Kępa and Rytel in a number of investigations over the years [39-41]. A comprehensive analysis and deperturbation calculations were conducted by Field et al. [42,43]. Later, Le Floch et al. [44] carried out a detailed study of perturbations appearing in the lowest $A^1\Pi$ ($\upsilon$ = 0) vibrational level. Perturbations occurring in the $A^1\Pi$ ($\upsilon$ = 0 - 4) levels were analyzed by Le Floch [45,46], who also derived very precise values of the terms for the $A^1\Pi$, $\upsilon$ = 0 - 8 levels. Recently, the lowest vibrational levels, $\upsilon$ = 0 - 4, of the $A^1\Pi$ state were deperturbed by Niu et al. [47,48] by means of accurate two-photon Doppler-free excitation using narrow band lasers with relative accuracy up to $\Delta\lambda/\lambda = 2 \times 10^{-8}$, as well as by a vacuum ultraviolet Fourier-transform spectrometer (VUV-FTS) installed on the DESIRS beamline at the SOLEIL synchrotron.

Deperturbation analyses of the $A^1\Pi$ state, in other CO isotopologues, were conducted by: Haridass et al. [49], Gavilan et al. [50], and Niu et al. [51] in $^{13}C^{16}O$; Haridass et al. [52,53] in $^{12}C^{18}O$ and $^{13}C^{18}O$; Beaty et al. [54] in $^{12}C^{18}O$; Hakalla et al. [55] in $^{12}C^{17}O$ by means of high-accuracy dispersive optical spectroscopy ($\Delta\lambda/\lambda$ up to about $1 \times 10^{-7}$) and VUV Fourier-transform measurements. For the less-abundant $^{13}C^{17}O$ isotopologue, the deperturbation analysis of the $A^1\Pi$ state has never been performed.

The $B^1\Sigma^+$ state is the first member of the *nsσ* Rydberg series of CO. The rotational perturbations occurring in the $B^1\Sigma^+$ ($\upsilon$ = 0) level were studied by Amiot et al. [56] in the infrared $C^1\Sigma^+$ - $B^1\Sigma^+$ spectra of the $^{12}C^{16}O$ and $^{14}C^{16}O$ isotopologues. Observed level-shift irregularities were very weak: 0.013 - 0.017 $cm^{-1}$ for $J$ = 4 and 6 in $^{12}C^{16}O$ as well as 0.003 - 0.005 $cm^{-1}$ for $J$ = 18 and 20 in $^{14}C^{16}O$, while the measurement accuracy was about 0.005 $cm^{-1}$. Janjić et al. [57] found a 'well marked' perturbation in the $B^1\Sigma^+$ ($\upsilon$ = 1, $J$ = 7 - 8) of $^{13}C^{16}O$, without assigning the perturber state(s).

This work focuses on a comprehensive deperturbation analysis of the $A^1\Pi$ ($\upsilon$ = 0, 1, 2, and 3) rovibrational levels in the $^{13}C^{17}O$ isotopologue. In order to perform this analysis, information on the following bands is included: the $^{13}C^{17}O$ $B^1\Sigma^+ \rightarrow A^1\Pi$ (0, 0), (0, 3), $B^1\Sigma^+ \leftarrow X^1\Sigma^+$ (0, 0), (1, 0), and $C^1\Sigma^+ \leftarrow X^1\Sigma^+$ (0, 0) [58], as well as previously published studies of the $B^1\Sigma^+ \rightarrow A^1\Pi$ (0, 1), (0, 2), (1, 0), (1, 1) [33,34] and $C^1\Sigma^+ \rightarrow A^1\Pi$ (0, 1), (0, 2), and (0, 3) [35]. The $B^1\Sigma^+ \leftarrow X^1\Sigma^+$, and $C^1\Sigma^+ \leftarrow X^1\Sigma^+$ systems are included in the study in order to determine level energies of $A^1\Pi$ in relation to the $X^1\Sigma^+$ ($\upsilon$ = 0, $J$ = 0) ground state of $^{13}C^{17}O$, to search for an occurrence of possible perturbations in the $B^1\Sigma^+$ ($\upsilon$ = 0) level and to introduce an independent set of improved constants for the $B^1\Sigma^+$ ($\upsilon$ = 0, 1) and $C^1\Sigma^+$ ($\upsilon$ = 0) levels. A comprehensive fit of the bands is performed to determine accurate deperturbed rotational constants and energy levels of the states under consideration. This paper is a sequel to work published earlier [47,48,51,55] and is leading to a follow-up with multi-isotope analysis.

## 2. Experiments

### *2.1. VIS-FTS data*

The $B^1\Sigma^+$ - $A^1\Pi$ (0, 0) and (0, 3) bands were measured in emission from a modified, water-cooled hollow-cathode discharge lamp with two anodes [59]. The lamp was initially filled with a mixture of helium and acetylene $^{13}C_2D_2$ (Cambridge Isotopes, 99.98% of $^{13}C$) at a pressure of approximately 7 Torr. A dc electric current was passed through the mixture for about 200 h. During that time, a small quantity of $^{13}C$ carbon became deposited on the inside of the cathode. Next, the lamp was evacuated, and then oxygen containing the $^{17}O_2$ isotope (Sigma-Aldrich, $^{17}O_2$ 70%) was admitted. The anodes were operated at 2 × 650 V and 2 × 50 mA dc in the conditions of a static gas pressure of 2 Torr inside the lamp. During the discharge process, the $^{17}O_2$ molecules react with the $^{13}C$-carbon atoms ejected from the cathode, thus forming the $^{13}C^{17}O$ molecules in the gas phase [34,35]. The temperature of the dc-plasma formed at the center of the cathode was about 650 K.

The high-resolution (HR) VIS emission spectra of the $^{13}C^{17}O$ $B^1\Sigma^+ \rightarrow A^1\Pi$ (0, 0) and (0, 3) bands were observed in the Materials Spectroscopy Laboratory, University of Rzeszów, with a 1.71-m Bruker IFS 125HR Fourier-transform spectrometer. The FTS was equipped with a quartz beamsplitter and a photomultiplier tube with housing and preamplifier, running in integration mode. The optical bench of the FTS was evacuated to a pressure lower than 0.002 hPa. The emission of the external CO source was focused onto the 1.0 mm diameter of the entrance aperture (field stop) of the FTS by a plano-convex quartz lens. In practice, the aperture is chosen to select as much as possible of the central region of the pattern of the interferogram. The permanently aligned interferometer has a 30° angle-of-incidence. The narrow angle-of-incidence makes more effective use of the beamsplitter by reducing beam polarization effects. The maximum optical path difference (OPD) of the FTS is 258 cm. The interferogram was obtained while recording the interferometric signal versus the OPD in a forward-backward acquisition mode. A 1.2 mW frequency-stabilized, single mode He-Ne laser controlled the position of the moving interferometer mirror (scanner), as well as provided a convenient internal wavenumber standard. The relative frequency stability of the laser beam was $\pm 2 \times 10^{-9}$ over 1 h.

The spectrum in the 11,000 - 25,000 $cm^{-1}$ wavenumber region was recorded at an instrumental resolution of 0.018 $cm^{-1}$ by coadding 128 scans in about 2.5 h of integration. Higher resolution was not necessary, since the full width at half maximum (FWHM) Doppler width of the CO lines is about 0.06 $cm^{-1}$ for the medium-strength, and single lines. The resolving power of the instrument was better than $10^6$. The recorded spectrum had a signal-to-noise ratio (SNR) of 70:1 and 100:1 for the $^{13}C^{17}O$ $B^1\Sigma^+ \rightarrow A^1\Pi$ (0, 0) and the (0, 3) band, respectively.

The spectrum was calibrated using the He-Ne 633 nm line (0.002 $cm^{-1}$ FWHM, 3000:1 SNR) of the internal frequency-stabilized laser. The line positions were measured by fitting Voigt lineshape functions to the experimental lines in a Levenberg-Marquardt procedure [60,61] included in the Bruker OPUS™ software package [62]. The estimated absolute 1$\sigma$ calibration uncertainty was 0.003 $cm^{-1}$. Finally, we expect the absolute accuracy of the line frequency measurements to be about 0.003 $cm^{-1}$ for single, medium-strength lines and about 0.01-0.02 $cm^{-1}$ for weak/blended ones.

While determining values of the molecular wavenumbers, blending due to spectral lines associated with other isotopologues, $^{13}C^{16}O$ B → A (0, 0) [51] and B → A (0, 3) [63], $^{12}C^{17}O$ B → A (0, 0) (theoretically evaluated) and B → A (0, 3) [55], as well as $^{12}C^{16}O$ B → A (0, 0), and (0, 3) [64] bands was taken into account. In the studied regions 17,950 - 18,650 $cm^{-1}$ and 22,150 - 22,500 $cm^{-1}$, in total 315 molecular emission lines of $^{13}C^{17}O$ were analyzed, 188 of which were interpreted as belonging to the (0, 0) and (0, 3) bands of the $B^1\Sigma^+ \rightarrow A^1\Pi$ system, and 122 were identified as lines deriving from the $B^1\Sigma^+ \rightarrow d^3\Delta_i$, $B^1\Sigma^+ \rightarrow e^3\Sigma^-$, $B^1\Sigma^+ \rightarrow a'^3\Sigma^+$, $B^1\Sigma^+ \rightarrow I^1\Sigma^-$, $B^1\Sigma^+ \rightarrow D^1\Delta$, $C^1\Sigma^+ \rightarrow e^3\Sigma^-$, $C^1\Sigma^+ \rightarrow a'^3\Sigma^+$, and $C^1\Sigma^+ \rightarrow D^1\Delta$ transitions. These 122 "extra-lines" refer to the spectral emission lines terminating on perturber states and gaining intensity from mixing with the $A^1\Pi$ state. Transition frequencies for the $^{13}C^{17}O$ $B^1\Sigma^+ \rightarrow A^1\Pi$ (0, 0) and (0, 3) bands are listed in Table 1, and wavenumbers connected with perturber states are listed in Table 2. During the deperturbation analysis, we extended and corrected the assignment of some heavily perturbed or weak $^{13}C^{17}O$ lines, previously analyzed [33-35]. They are collected in Table 3. An overview of the observed $^{13}C^{17}O$ $B^1\Sigma^+$ ($\upsilon$ = 0) → $A^1\Pi$ ($\upsilon$ = 0, 3) spectra, together with extra-lines, rotational assignments and simulated spectra are shown in Figs. 1 and 2. The ratios of the molecular gas compositions used to obtain the spectra were $^{13}C^{17}O$ : $^{13}C^{16}O$ : $^{12}C^{17}O$ : $^{12}C^{16}O$ = 1 : 0.76 : 0.34 : 0.25.

*2.2. VUV-FTS data*

VUV spectra are recorded with the Fourier-transform spectrometer installed on the DESIRS beamline at the SOLEIL synchrotron. This spectrometer is described in great detail in [65-68]. $^{13}C^{17}O$ $A^1\Pi \leftarrow X^1\Sigma^+$ photoabsorption bands have been recently observed for the first time by Lemaire et al. [32] in the course of a $^{13}C^{18}O$ study. $^{13}C^{17}O$ bands have been also observed in their $^{13}C^{16}O$ sample. The occurrence of these $^{13}C^{17}O$ bands is due to < 0.1% isotopic contamination of $^{13}C^{17}O$ in samples of $^{13}C^{16}O$ and $^{13}C^{18}O$, and are accompanied by much stronger absorption from the main isotopologues. Full details including transition frequencies, term values and molecular constants of the multi-isotopologue measurements of $^{13}C^{17}O$ $B^1\Sigma^+ \leftarrow X^1\Sigma^+$ (0, 0), $B^1\Sigma^+ \leftarrow X^1\Sigma^+$ (1, 0) and $C^1\Sigma^+ \leftarrow X^1\Sigma^+$ (0, 0) and other bands will be published in a separate paper [58]. Figures 3, 4, and 5 show the corresponding $^{13}C^{17}O$ spectra obtained after subtracting from a $^{13}C^{16}O$ raw spectrum a model including $^{12}C^{16}O$, $^{13}C^{16}O$, and $^{13}C^{18}O$. At the high pressures used to record spectra with nominally $^{13}C^{16}O$ or $^{13}C^{18}O$ gas species, lines of the dominant gas are strongly saturated, and lines from the trace isotopologues, particularly $^{13}C^{17}O$, can be identified. Based on the relative strengths of the absorption features, the ratio of the isotopologues in the $^{13}C^{16}O$ gas sample, sorted by concentration, was found to be $^{13}C^{16}O$ : $^{13}C^{18}O$ : $^{13}C^{17}O$ : $^{12}C^{16}O$ = 1 : 0.041 : 0.073 : 0.0045. Relevant to this study are the $J$-range of $^{13}C^{17}O$ rotational energy levels and their accuracy deduced from these spectra. We note that the $^{13}C^{17}O$ results extracted from $^{13}C^{16}O$ and $^{13}C^{18}O$ transition frequencies are less accurately determined close to the band origin (i.e. at low $J'$ values) due to the saturation effects inducing blending: $J' < 2$ for B($\upsilon$=0), $J' < 7$ for C($\upsilon$=0). This effect becomes much weaker or absent for the high $J'$ of B($\upsilon$=0) and C($\upsilon$=0), and all $J'$ values of B($\upsilon$=1) as the lines of the different isotopologues become more separated while $\upsilon'$ and $J'$ increase and the saturation of the dominant species disappears. Thus the results for low $J'$ can be reliably extrapolated from the observed levels (up to $J$ = 23 for B($\upsilon$=0), 21 for B($\upsilon$=1) and 39 for C($\upsilon$=0)). The total uncertainty of the VUV-FTS-measured B($\upsilon$=0), B($\upsilon$=1), and C($\upsilon$=0) level energies is 0.04 $cm^{-1}$ on average and always lower than 0.1 $cm^{-1}$. These uncertainties are a combination of statistical fitting errors (different for every line, estimated to fall between 0.001 and 0.09 $cm^{-1}$) and calibration error of the wavenumber scale (estimated to be 0.01 $cm^{-1}$). The wavenumber calibration was made separately for each band with reference to various CO absorption lines measured in the highly-accurate laser-based work of Drabbels et al. [69], Ubachs et al. [30], and Cacciani et al. [70].

## 3. Results

### *3.1. Perturbations in CO*

The CO molecular orbitals are conventionally labeled in energy and symmetry order: 1σ (σ1s), 2σ (σ*1s), 3σ (σ2s), 4σ (σ*2s), 1π (π2p), 5σ (σ2p), 2π (π*2p), 6σ (3sσ), 7σ (3pσ), where (1 - 2)σ are core orbitals, (3 - 5)σ and (1 - 2)π are valence orbitals, and (6 - 7)σ are Rydberg orbitals. The principal configurations of the electronic states considered in this paper are:

$X^1\Sigma^+$ : $1\sigma^2\, 2\sigma^2\, 3\sigma^2\, 4\sigma^2\, 1\pi^4\, 5\sigma^2$,

$A^1\Pi$ and $a^3\Pi$ : $1\sigma^2\, 2\sigma^2\, 3\sigma^2\, 4\sigma^2\, 1\pi^4\, 5\sigma^1\, 2\pi^1$

$a'^3\Sigma^+$, $e^3\Sigma^-$, $d^3\Delta_i$, $I^1\Sigma^-$, $D^1\Delta$ : $1\sigma^2\, 2\sigma^2\, 3\sigma^2\, 4\sigma^2\, 1\pi^3\, 5\sigma^2\, 2\pi^1$

$B^1\Sigma^+$ or $C^1\Sigma^+$ : $1\sigma^2\, 2\sigma^2\, 3\sigma^2\, 4\sigma^2\, 1\pi^4\, 5\sigma^1$ ($6\sigma^1$ or $7\sigma^1$)

The spin-orbit and $L$-uncoupling perturbations between the $A^1\Pi$ state and the $a'^3\Sigma^+$, $e^3\Sigma^-$, $d^3\Delta_i$, and $I^1\Sigma^-$, $D^1\Delta$ states, respectively, involve a $1\pi \rightarrow 5\sigma$ orbital promotion. The possible $L$-uncoupling perturbation between the $A^1\Pi$ state and the $B^1\Sigma^+$ state involves a $2\pi \rightarrow 6\sigma$ orbital promotion. The magnitude of this $2\pi \rightarrow 6\sigma$ perturbation matrix element is likely to be weaker than that for the $L$-uncoupling $1\pi \rightarrow 5\sigma$ perturbations because the $1\pi$ and 5σ orbitals are mostly of O-atom localized 2p character while the $2\pi$ orbital resembles a dδ Rydberg orbital and the 6σ orbital is an sσ Rydberg orbital. The shape of the $2\pi$ molecular antibonding valence orbital resembles that of an atomic Rydberg orbital with orbital angular momentum $2\hbar$ and projection of angular momentum along the molecular axis of $1\hbar$: |2,1>. The possible perturbation between the $e^3\Sigma^-$ and $B^1\Sigma^+$ states involves a $(2\pi, 5\sigma) \rightarrow (1\pi, 6\sigma)$ double-promotion. Since the spin-orbit operator is a one-electron operator, which has a Δ(s-o) = 0 or 1 change in spin-orbital (s-o) selection rule, the e ~ B spin-orbit perturbation matrix element is zero in first-order. A second-order interaction, mediated by a nearby vibrational level of the $a^3\Pi$ or $A^1\Pi$ state, could result in a very weak e ~ a/A ~ B indirect perturbation. The possibility of a perturbation of the $B^1\Sigma^+$ state by *direct* spin-orbit interaction with the $a'^3\Sigma^+$ state can be ruled out by the $\Delta\Omega = 0$ and $\pm \leftrightarrow \pm$ parity selection rule for spin-orbit perturbations. The $a'^3\Sigma^+$ state contains only an $0^-$ component and the $B^1\Sigma^+$ state consists exclusively of an $0^+$ component. It is in principle possible to have an *indirect* interaction between the a' and B states "*mediated* by" the $\Omega = 1$ component of an intermediate $a^3\Pi$ or $A^1\Pi$ state:

$$\langle\, {}^3\Sigma_1^+ | \boldsymbol{H^{SO}} |\, {}^3\Pi_1 \rangle \langle\, {}^3\Pi_1\, | \boldsymbol{H^{rot}}(S - uncoupling) |\, {}^3\Pi_0 \rangle \langle\, {}^3\Pi_0\, | \boldsymbol{H^{SO}} |\, {}^1\Sigma_0^+ \rangle, \tag{1}$$

or

$$\langle\, {}^3\Sigma_1^+ | \boldsymbol{H^{SO}} |\, {}^1\Pi_1 \rangle \langle\, {}^1\Pi_1\, | \boldsymbol{H^{rot}}(L - uncoupling) |\, {}^1\Sigma_0^+ \rangle. \tag{2}$$

These two perturbation paths can be ruled out by the small computed values of the vibrational overlap integrals between near-degenerate vibrational levels of the B and A or a electronic states, $\langle v_B | v_A \rangle$ and $\langle v_B | v_a \rangle$. The possibility of a direct spin-orbit interaction between the $B^1\Sigma^+$ state and a near-degenerate vibrational level of the $a^3\Pi_{0+}$ state can be ruled out by the small value of the computed $\langle v_B | v_a \rangle$ vibrational overlap integral.

It is always possible to evaluate a matrix element using basis functions expressed in any Hund's coupling case. For many reasons, we find case (a) to be most convenient for valence states of almost all diatomic molecules. The spin-orbit operator is diagonal in Ω in Hund's case (a). The spin-orbit matrix element between pure Ω = 0 and 1 states is rigorously zero. This is true for all spin and all Λ basis states. All terms in the field free Hamiltonian are diagonal in parity. The eigenstates of a $^3\Sigma^+$ state for each value of *J* (*J* is rigorously conserved at zero electric field; nonzero nuclear spins can weakly break the *J* quantum number, but this effect is negligible for $^{13}C^{17}O$) are linear combinations of Ω = 0 and 1 for *f*-symmetry and pure Ω = 1 for *e*-symmetry. The Ω = 0 basis state of $^3\Sigma^+$ has *f* symmetry but the $^1\Sigma^+$ state consists exclusively of *e*-symmetry Ω = 0 basis states. Hence, there can be no perturbation of $^1\Sigma^+$ by a $^3\Sigma^+$ state. Contrary to the opinion of Kovács [71] (Fig. 4.55 p. 267), the $F_2$ component of a $^3\Sigma^+$ state is a pure Ω = 1 state. It cannot perturb $^1\Sigma^+$ via spin-orbit because that is a pure Ω = 0 state. In order for such a perturbation to occur, there needs to be a third electronic state, either $^3\Pi$ or $^1\Pi$, that can perturb via a combination of *L*-uncoupling and spin-orbit mechanisms. Also, rotational interactions between rovibronic levels of the $^1\Sigma^+$ state and levels of $^3\Delta$, $^1\Delta$, and $^1\Sigma^-$ states are forbidden by the ΔS = 0, ΔΩ = 0, ±1, and ± ↔ ± parity selection rules for rotational (L.J "*L*-uncoupling" and L.S "spin-electronic") interaction terms [36,72].

*3.2. Irregularities of the $^{13}C^{17}O$ $A^1\Pi$ and $B^1\Sigma^+$ states*

Predicted maxima of perturbations and identification of responsible perturbers result from the rovibronic term-crossing diagrams of the $^{13}C^{17}O$ $A^1\Pi$ ($v$ = 0 - 3), $d^3\Delta$ ($v$ = 4 - 9), $e^3\Sigma^-$ ($v$ = 1 - 6), $a'^3\Sigma^+$ ($v$ = 9 - 14), $I^1\Sigma^-$ ($v$ = 0 - 5), and $D^1\Delta$ ($v$ = 0 - 4) levels, which were presented in our previous works [33-35]. A term-crossing diagram of the $^{13}C^{17}O$ $B^1\Sigma^+$ ($v$ = 0) level containing irregularities and close lying $A^1\Pi$ ($v$ = 20 - 22), $e^3\Sigma^-$ ($v$ = 29 - 38), $a'^3\Sigma^+$ ($v$ = 39 - 43), and $a^3\Pi$ ($v$ = 31 - 33) levels is shown in Fig. 6. The diagram was plotted on the basis of $^{13}C^{17}O$ equilibrium molecular constants given by Hakalla et al. [34] for the $A^1\Pi$ state, isotopically recalculated equilibrium molecular constants given by Field [42] for the $d^3\Delta_i$, $e^3\Sigma^-$, $a'^3\Sigma^+$, and $I^1\Sigma^-$ states, by Havenith et al. [73] for the $a^3\Pi$ state, as well as by Kittrell et al. [74] for the $D^1\Delta$ state. The $T_e$ values were taken from Refs. [74-76] and the $G(v = 0)$ value of the $^{13}C^{17}O$ $X^1\Sigma^+$ ground state comes from Coxon et al. [77]. Note that in some cases we will deal with observable effects of interactions that originate from some perturbers that do not intersect with the $A^1\Pi$ term curves.

Thanks to high accuracy VIS and VUV FTS measurements carried out in the $^{13}C^{17}O$ isotopologue, weak (0.05 - 0.75 $cm^{-1}$) irregularities appearing in the $B^1\Sigma^+$ ($v$ = 0) level at $J$ = 0, 7, 16, 21, 26, and 34 - 36 were noticed. However, identification of the states responsible for these possible perturbations is complicated due to indirect interaction problems discussed in the previous subsection. We were not able to determine the dependence between the effective perturbation parameters and isotopologue-independent electronic factors ***a*** and ***b*** [55] for the $B^1\Sigma^+$ ~ perturbers' interactions, because relevant single-orbital perturbation matrix element values were too small to extract perturbation parameters. In addition, the studied region is close to the ($^3P_0$ + $^3P_2$) first dissociation limit of the CO molecule (90,679 $cm^{-1}$ [78]), where many singlet and triplet states converge, forming a region with very densely spaced vibrational levels, which complicates the identification of a perturber. Moreover, the vibrational overlap integrals involving the $B^1\Sigma^+$ state and the $A^1\Pi$ and/or $a^3\Pi$ states as mediating ones are very small, at about $1 \times 10^{-3}$ - $1 \times 10^{-5}$. Therefore, an identification of the $B^1\Sigma^+$ perturbers was not possible. Thus, the transition frequencies which involved the unidentified perturbers were appropriately weighted in the final deperturbation fit, so that they would have a negligible influence on the quality of the results.

### *3.3. Deperturbation analysis*

We model each observed $^{13}C^{17}O$ $B^1\Sigma^+$ ($\upsilon$ = 0, 1) → $A^1\Pi$ ($\upsilon$ = 0 - 3) [33,34], $C^1\Sigma^+$ ($\upsilon$ = 0) → $A^1\Pi$ ($\upsilon$ = 1 - 3) [35], $B^1\Sigma^+$ ($\upsilon$ = 0, 1) ← $X^1\Sigma^+$ ($\upsilon$ = 0), and $C^1\Sigma^+$ ($\upsilon$ = 0) ← $X^1\Sigma^+$ ($\upsilon$ = 0) band and interacting levels based on a local deperturbation analysis, similar as was done by Hakalla et al. [55]. In total, 1003 transitions from 12 bands of $^{13}C^{17}O$ were used in the procedure of disentangling $A^1\Pi$ from the $d^3\Delta$, $e^3\Sigma^-$, $a'^3\Sigma^+$, $I^1\Sigma^-$, and $D^1\Delta$ states using the Pgopher software [79]. This procedure results in 63 deperturbed molecular parameters fitted for the less-abundant $^{13}C^{17}O$ isotopologue. The computed level positions, line wavenumbers and intensities are the result of a matrix diagonalization taking into account all interacting levels. The assignment of $A^1\Pi(\upsilon)$ and perturber levels, the selection of which interactions and parameters could be determined from the spectra, and their values, were iteratively optimized. The Pgopher program [79] is based on an effective Hamiltonian with matrix elements similar to Field [42], Bergeman et al. [80], and Le Floch et al. [44], as we present in the Supplementary Material, retaining their symbols for the various molecular constants. The Hamiltonian matrix used in the present deperturbation analysis contains not only the perturber levels which cross the $A^1\Pi(\upsilon)$ level, but also those which interact with it without intersections. The wavenumbers of intense, moderately strong and isolated spectral lines were assigned relative weights of 1.0 during the fitting. However, the frequencies of the weak and/or blended lines are less accurate, so they were individually weighted between 0.5 - 0.1, depending on their weakness and/or extent of blending as well as failure to recognize the causes of irregularities in B ($\upsilon$ = 0).

The homogeneous interactions of the short-lived $A^1\Pi$ singlet levels with the long-lived $e^3\Sigma^-$, $a'^3\Sigma^+$, and $d^3\Delta_i$ triplet levels result from the spin-orbit term coupling, represented by *J*-independent matrix elements. In turn, the interactions of $A^1\Pi$ with $I^1\Sigma^-$ and $D^1\Delta$ are caused by *L*-uncoupling and, therefore, produce heterogeneous interactions with *J*-dependent matrix elements [36]. The interaction parameters are denoted by $\eta_{v,v'}$ and $\xi_{v,v'}$ for the spin-orbit and *L*-uncoupling perturbations, respectively.

For some of the levels under consideration, there are insufficient term-value data to determine molecular constants independently of the deperturbation calculation. In such cases, we adopted isotopically recalculated molecular constants of these $d^3\Delta$, $e^3\Sigma^-$, $a'^3\Sigma^+$, $I^1\Sigma^-$, and $D^1\Delta$ states, using Dunham's relationship within the Born-Oppenheimer approximation [81] from the data given in Refs. [42,74]. These values were held fixed during the fitting procedures. In the calculations we included all possible perturber levels, listed in Table 4, which have a non-negligible influence on the $^{13}C^{17}O$ $A^1\Pi$ ($v$ = 0-3), even if some of them do not exhibit crossings, as it is possible to see in Fig. 7. Imperfections in our treatment of this non-negligible influence lead to a root-mean-square error (rmse) of the model fit to affected lines which exceeds their weighted experimental uncertainties.

The rmse value~~s~~ of the unweighted residuals of the wavenumbers used in the deperturbation fit are dominated by the uncertainties of the weakest B-A (1, 0) and (1, 1) bands obtained in Ref. [34]. The rmse~~average~~ value of the weighted residuals of the ~~weighted~~ wavenumbers used in the deperturbation fit based on the data from~~of~~ VIS-VUV FTS, is 0.007 cm$^{-1}$, so the fitting model acceptably reproduces the experimental data set.

In some cases, it was impossible to calculate the $\eta_{v,v'}$ and $\xi_{v,v'}$ interaction parameters, because their fitting was statistically unjustified. This occurs when we deal with an insufficient quantity of experimental transitions attributed to perturber states in the neighborhood of the avoided crossings or when interacting states are too distant. The solution to this problem was a semi-empirical method of estimation of the missing interaction parameters provided in Refs. [36,42-44]. It shows that the perturbation matrix element ($\alpha$, $\beta$) is the product of a vibrational factor and an electronic perturbation parameter (***a***, ***b***). The appropriate dependences are given in the Supplementary Material. The isotopologue-independent parameters ***a*** and ***b*** are characteristic of the relevant electronic configuration within the single-configuration approximation and they are strongly dependent on the internuclear separation of the molecule [82]. An approach to calculate of $\eta_{v,v'}$ and $\xi_{v,v'}$ for the $A^1\Pi(v)$ state and its interacting levels is described in our previous work [55]. The isotopologue-independent purely electronic perturbation parameters $\boldsymbol{a}_{A\sim d,e,a'}$ and $\boldsymbol{b}_{A\sim I,D}$ were taken from Le Floch [44]. The $\langle v_A | v_{d,e,a'} \rangle$ vibrational overlap integrals, as well as the $\langle v_A | \boldsymbol{B}(\boldsymbol{R}) | v_{I,D} \rangle$ rotational operator integrals of $^{13}C^{17}O$, were calculated on the basis of RKR parameters for the $^{13}C^{17}O$ A, d, e, a', I, and D states. The RKR values were obtained from isotopically recalculated equilibrium constants of Field [42], Field et al. [36,43], Le Floch et al. [44], Kittrell et al. [74] and Hakalla et al. [34] using the computer program 'FRACONB' of Jung [83] and Jakubek [84]. Finally, the missing perturbation parameters under consideration were estimated and held fixed during the fitting.

As expected from the extensive coverage of the $^{13}C^{17}O$ transition energies in the 12 bands, we fitted accurate deperturbed molecular constants for the $A^1\Pi$ ($\upsilon$ = 0, 1, 2, and 3) levels and their perturbers, where the definitions of the constants are the same as in Ref. [47,48,55]. The results are collected in Table 4. We also obtained the molecular constants for the $B^1\Sigma^+$ ($\upsilon$ = 0, 1) and $C^1\Sigma^+$ ($\upsilon$ = 0) levels, results of which are given in Table 5. The constants of the $B^1\Sigma^+$ ($\upsilon$ = 0, 1) and $C^1\Sigma^+$ ($\upsilon$ = 0) levels are compared with analogous values derived in previous studies [33-35]. The correlation matrix was checked carefully and it was found that it shows satisfactorily low correlations between fitted parameters of the model.

All perturbation parameters that are the result of the global deperturbation analysis of the $^{13}C^{17}O$ isotopologue are listed in Table 6. The electronic part of each perturbation parameter was obtained by dividing the spin-orbit parameter $\eta_{\upsilon,\upsilon'}$ by the vibrational overlap $\langle\upsilon|\upsilon'\rangle$, or by dividing the rotation-electronic parameter $\xi_{\upsilon,\upsilon'}$ by the rotational operator integral $\langle\upsilon|\boldsymbol{B}(r)|\upsilon'\rangle$. Table 6 also includes *r*-centroids, which are indicators of the internuclear distance of maximum vibrational overlap. It may be noted that for perturbations between any pair of electronic states, the *r*-centroids are approximately constant. In this table, for comparison, we also included the results of all the tests to determine the isotopologue-independent electronic perturbation parameters ***a*** and ***b***, which have been carried out so far, that is, by Field et al. [43], Le Floch [44], Haridass et al. [52], and Hakalla et al. [55].

### 3.4. *Level energies*

The rotational terms of the $^{13}C^{17}O$ $A^1\Pi$ ($\upsilon$ = 0, 1, 2 and 3) vibrational levels referenced to the $X^1\Sigma^+$ ($\upsilon$ = 0, *J* = 0) ground state were calculated as differences of the $B^1\Sigma^+$ ($\upsilon$ = 0, 1), $C^1\Sigma^+$ ($\upsilon$ = 0) term values obtained from a VUV-FTS experiment [58] and B → A (0 - $\upsilon$") (this work and [33,34]), C → A (0 - $\upsilon$") [35] transition frequencies. The $^{13}C^{17}O$ ($B^1\Sigma^+$, $C^1\Sigma^+$) → ($d^3\Delta_i$, $e^3\Sigma^-$, $a'^3\Sigma^+$, $I^1\Sigma^-$, $D^1\Delta$) extra-lines, listed in Table 2, were used to determine level energies of the D, I, e, a', and d states in $^{13}C^{17}O$. The higher *J*-terms of the $^{13}C^{17}O$ $A^1\Pi$ ($\upsilon$ = 0 - 3) levels were calculated using deperturbed $T_\upsilon$ constants from Table 4 and relative level energies of the $A^1\Pi$ state obtained from B → A (0 - $\upsilon$") [33,34], C → A (0 - $\upsilon$") [35] bands by means of the linear least-squares method in the version given by Curl and Dane [85] and Watson [86]. The final term values were merged by using the weighted average method. The values for the $A^1\Pi$ ($\upsilon$ = 0, 1, 2 and 3) terms are collected in Table 7 and for the d($\upsilon$), e($\upsilon$), a'($\upsilon$), D($\upsilon$), and I($\upsilon$) levels in Table 8. The results for $B^1\Sigma^+$ ($\upsilon$ = 0, 1), $C^1\Sigma^+$ ($\upsilon$ = 0), and other levels in the 115-101 nm range will appear in a forthcoming publication [58].

The reduced term values $T(J) - BJ(J+1) + DJ^2(J+1)^2$ of the perturbed $^{13}C^{17}O$ $A^1\Pi$ ($\upsilon$ = 0 - 3) and $B^1\Sigma^+$ ($\upsilon$ = 0) levels, in relation to the $\upsilon$ = 0, *J* = 0 level of the $X^1\Sigma^+$ ground state, together with the hypothetical unperturbed and crossing perturber levels, are presented in Fig. 7. These reduced terms of perturbers, which could not be derived from our current analysis, were calculated on the basis of isotopically recalculated equilibrium molecular constants given by Field [42] for $d^3\Delta_i$, $e^3\Sigma^-$, $a'^3\Sigma^+$, and $I^1\Sigma^-$ states, by Kittrell et al. [74] for $D^1\Delta$ state as well as by Havenith et al. [73] for the $a^3\Pi$ state. $T_e$ values were taken from Refs. [74-76] and the $G$ ($\upsilon$ = 0) value of $X^1\Sigma^+$ from Coxon et al. [77].

It should be noted that one could make a distinction between accuracy and precision. The first referring to the absolute wavenumber scale of importance for deriving term energies with respect to X ($\upsilon$ = 0, $J$ = 0) state (not better than the 0.04 cm$^{-1}$) and the other refers to the fitting with respect to the relative scale (at about 0.003 cm$^{-1}$). The quality of the deperturbation analysis reflects the relative precision (0.003 cm$^{-1}$) of the measurements.

## 4. Discussion

The high accuracy VIS-VUV FTS measurements carried out on the $^{13}C^{17}O$ isotopologue reveal very weak irregularities appearing in the $B^1\Sigma^+$ ($\upsilon$ = 0) level and contribute to a successful global deperturbation fit. As a result accurate deperturbed molecular constants, term values, spin-orbit and $L$-uncoupling interaction parameters as well as isotopologue-independent spin-orbit and rotation-electronic perturbation parameters of the $^{13}C^{17}O$ $A^1\Pi$ ($\upsilon$ = 0, 1, 2 and 3) state could be derived. The analysis allowed also for verification and improvement of the perturbations in the $^{13}C^{17}O$ $A^1\Pi$ ($\upsilon$ = 0 - 3) levels reported in Ref. [33-35]. Comparing Table 4 with the data given in the mentioned publications, we can see that the $I^1\Sigma^-$ ($\upsilon$ = 0), $d^3\Delta_i$ ($\upsilon$ = 5), and $e^3\Sigma^-$ ($\upsilon$ = 5) rovibronic levels have a weak impact on the $\upsilon$ = 0, $\upsilon$ = 1, and $\upsilon$ = 3 levels of the $A^1\Pi$ state, respectively. The reduced terms presented in Fig. 7 indicate that the rotational progressions of these perturbers do not cross with the $A^1\Pi$ ($\upsilon$ = 0, 1, and 3) levels, and as a result they cause global energy shifts of the $A^1\Pi$ ($\upsilon$) levels. This phenomenon is precisely the same as a vibrational perturbation shifting an entire band systematically. Also, all $J$'s of maximum irregularities of the $A^1\Pi$ ($\upsilon$ = 0, 1, and 3) levels, which have been observed so far [33-35], were verified.

We tried to find the alleged perturbers responsible for irregularities appearing in the $B^1\Sigma^+$ ($\upsilon$ = 0) level. As shown in Fig. 7a, a possible perturber of B ($\upsilon$ = 0, $J$ = 7) level could be the $e^3\Sigma^-$ ($\upsilon$ = 29) state, but this would require a second-order interaction beyond the single-configuration picture of diabatic electronic states, or else a mediating interaction with a close lying $a^3\Pi$ or $A^1\Pi$ state, i.e. a(31) or A(19 or 20). However, *i*) the a(31), A(19), and A(20) levels lie too far from e(29), *ii*) the $^{13}C^{17}O$ $\langle \upsilon_{B(0)} | \boldsymbol{B(R)} | \upsilon_{a(31)} \rangle$, $\langle \upsilon_{B(0)} | \boldsymbol{B(R)} | \upsilon_{A(19)} \rangle$, and $\langle \upsilon_{B(0)} | \boldsymbol{B(R)} | \upsilon_{A(20)} \rangle$ rotational operator integrals equal -9.3 × 10$^{-5}$ cm$^{-1}$, -5.8 × 10$^{-3}$ cm$^{-1}$, and 4.2 × 10$^{-3}$ cm$^{-1}$ , respectively, so they are too small to cause the expected effect. Similarly for B ($\upsilon$ = 0, $J$ = 21), where the A(20) state is too distant to cause the B(0) ~ A(20) ~ e(30) indirect perturbation and the $\langle \upsilon_{B(0)} | \boldsymbol{B(R)} | \upsilon_{A(20)} \rangle$ vibrational overlap integral is too small (-9.3 × 10$^{-5}$ cm$^{-1}$). The $J$ = 34 level of B(0) could be affected by an indirect B(0) ~ a(32) ~ e(32) perturbation, but $\langle \upsilon_{B(0)} | \boldsymbol{B}(\boldsymbol{R}) | \upsilon_{a(32)} \rangle$ = 7.7 × $10^{-5}$ $cm^{-1}$ is far too small. Finally, the $J$ = 35 level of B(0) could be directly perturbed via an $L$-uncoupling interaction with A(21), but the vibrational overlap integral, $\langle \upsilon_{B(0)} | \boldsymbol{B}(\boldsymbol{R}) | \upsilon_{A(20)} \rangle$ = -3.1 × $10^{-3}$ $cm^{-1}$, is also not big enough for this purpose. We note that the term values of the states suspected of perturbing the $B^1\Sigma^+$ ($\upsilon$ = 0) level were calculated on the basis of long extrapolations to high rovibrational levels using equilibrium molecular constants. This results in large uncertainties in the level positions determinations. For all these reasons we could not unequivocally find a cause for the alleged perturbations in the $B^1\Sigma^+$ ($\upsilon$ = 0) level. Therefore, the molecular constants obtained for this level, presented in Table 5, should be considered as effective ones. The only other possibility is that another hitherto unknown Rydberg state is involved, of $^1\Pi$, $^3\Pi$, or $^3\Sigma^-$ character. We hope, that our future work [87] concerning a multi-isotopologue analysis of the CO molecule may shed more light on this problem.

The $q_v$ constants in the CO $A^1\Pi$ state are very small as well as they have erratic signs and magnitudes (Table 4). There can be several reasons. The $q_v$ constants are purely a result of second-order perturbation theory and predominantly determined by the $I^1\Sigma^-$ state. In addition, the $\Lambda$-doubling of the $A^1\Pi$ state is strongly dependent on local perturbations, so that it is negligible away from the crossing regions. The thing that is remarkable about non-hydride molecules is that the $\Lambda$-doubling in $^1\Pi$ states is always very small [36,47,48]. The magnitude of $q_v$ is determined by a perturbation summation. No particular significance can be associated with its sign or magnitude. However, in order to reproduce the observed rotational levels, it is necessary to retain most of the fitted digits.

We should note that the perturbation parameters that control the interactions between the $d^3\Delta_i$, $e^3\Sigma^-$, $a'^3\Sigma^+$ states and the $A^1\Pi$ state are the cumulative result of the interactions of all three $\Omega$-components of triplet states. It is obvious, however, that each of the $F_1$, $F_2$, and $F_3$ components interacts with appropriate components of the perturbed state to a varying degree. It is clearly seen in Fig. 7b for the $A^1\Pi$ ($\upsilon$ = 0) ~ $d^3\Delta_i$ ($\upsilon$ = 4) interaction, where the second and third $\Omega$-component of the $d^3\Delta_i$ ($\upsilon$ = 4) state causes more severe deviations (about 4.5 $cm^{-1}$) of the $A^1\Pi$ ($\upsilon$ = 0) level than the first one (about 0.5 $cm^{-1}$). One of the causes could be the fact that strong spin-orbit interaction between the states in the Hund's case ($a$) requires $\Delta\Omega = 0$. Thus, the third component ($\Omega$ = 1) of the $d^3\Delta_i$ state should have the strongest influence on the $A^1\Pi$ ($\Omega$ = 1) state and that is what we have observed.

In Table 5 one can notice minor inconsistencies in the $B_\upsilon$ and $D_\upsilon$ rotational constants of the $B^1\Sigma^+$ ($\upsilon$ = 0, 1) and $C^1\Sigma^+$ ($\upsilon$ = 0) levels in relation to our previous results [33-35]. These minor discrepancies may result from: *i*) the first attempt of analysis of the weak perturbations occurring at $^{13}C^{17}O$ $B^1\Sigma^+$ ($\upsilon$ = 0) level, *ii*) the differences in calculation methods used to obtain the constants, as we described in Ref. [55], and/or *iii*) band measurements that include higher-$J$ levels, having a significant impact on the rotational distortion $D_\upsilon$ constants, always strongly correlated with both $B_\upsilon$ and $T_\upsilon$ constants.

Table 6 shows the perturbation parameters obtained in this work. The isotopic independence of the ***a*** spin-orbit and ***b*** rotation-electronic perturbation parameters for each state that interacts with the $A^1\Pi$ state is evident. The parameters also remain satisfactorily consistent with analogous values determined in the deperturbation calculations of $^{12}C^{16}O$ [43,44], $^{12}C^{18}O$ [52], and $^{12}C^{17}O$ [55].

## 5. Conclusion

Two high precision experimental methods, photoemission in a discharge with Fourier-transform spectroscopy in the visible region as well as synchrotron absorption FTS in the vacuum ultraviolet region, were used to obtain first high-resolution spectra of the $B^1\Sigma^+ \rightarrow A^1\Pi$ (0, 0), (0, 3), and the $B^1\Sigma^+ \leftarrow X^1\Sigma^+$ (0, 0), (1, 0), $C^1\Sigma^+ \leftarrow X^1\Sigma^+$ (0, 0) bands, respectively. Combining the new information and our previous results [33-35], in total 1003 rotational lines in 12 bands of $^{13}C^{17}O$ ($B^1\Sigma^+ \rightarrow A^1\Pi$, $C^1\Sigma^+ \rightarrow A^1\Pi$, $B^1\Sigma^+ \leftarrow X^1\Sigma^+$, $C^1\Sigma^+ \leftarrow X^1\Sigma^+$) and in 18 bands of extra-lines, and literature data on CO isotopologues, allowed for a global deperturbation analysis of this complex of interacting levels and the fitting of model energy levels to all experimental data within their uncertainties. 63 independent adjustable parameters were required for this fit. The model describes the $A^1\Pi$ ($\upsilon$ = 0 - 3), $B^1\Sigma^+$ ($\upsilon$ = 0, 1), $C^1\Sigma^+$($\upsilon$ = 0), $d^3\Delta$ ($\upsilon$ = 4, 5, 7, 8), $e^3\Sigma^-$ ($\upsilon$ = 1, 3, 4, 5), $a'^3\Sigma^+$ ($\upsilon$ = 9, 10, 12, 13), $I^1\Sigma^-$ ($\upsilon$ = 0, 2, 3, 5), and $D^1\Delta$ ($\upsilon$ = 1, 4) levels, as well as the spin-orbit and *L*-uncoupling parameters responsible for the interactions of the $A^1\Pi$ ($\upsilon$ = 0 - 3) levels with their perturbers. Isotopologue-independent perturbation parameters of the $A^1\Pi$ interactions were also derived. With these combined data, 335 rotational terms of $A^1\Pi$ ($\upsilon$ = 0 - 3) levels and their perturbers were determined up to $J$ = 36.

The CO A-state has been a case study for diatomic molecular deperturbation studies. The mass-dependent shifts of energy levels mean each isotopologue exhibits different perturbation effects, but each arises from the same underlying rotational and spin-orbit interactions. The present study of $^{13}C^{17}O$ provides an additional perspective on the non-Born-Oppenheimer and relativistic effects common to all isotopologues.

## Acknowledgements

R. Hakalla thanks LASERLAB-EUROPE for support of this research via funding Grants EUH2020-RIP-654148 and EC’s-SFP-284464. We are grateful to the general and technical staff of SOLEIL synchrotron as well as to the scientific committee for providing beam time under Project numbers 20110121, 20120653, and 20140051. R.W. Field thanks the US National Science Foundation [Grant number CHE-1361865] for support of his research, which includes substantive collaborations. J.R. Lyons and G. Stark thank the NASA Origins of Solar Systems program for support [Grant number NNX14AD49G]. S. Federman was supported by NASA Grants NNG 06-GG70G and NNX10AD80G to the University of Toledo. J.-L. Lemaire thanks the ISMO-CNRS (Institut des Sciences Moléculaires d'Orsay at Université Paris-Sud) for his welcome as honorary professor. R. Hakalla would like to express his gratitude for the support of the European Regional Development Fund and the Polish state budget within the framework of the Carpathian Regional Operational Programme [Grant RPPK.01.03.00-18-001/10] for the period of 2007–2013 through the funding of the Centre for Innovation and Transfer of Natural Science and Engineering Knowledge of the University of Rzeszów.

The Supplementary Material is deposited in the Electronic Depository of the Supplementary Material of the journal.

## Table 1

$B^1\Sigma^+ \rightarrow A^1\Pi$ (0, 0) and (0, 3) transitions identified in the VIS-FT spectra of $^{13}C^{17}O$.[a, b, c]

| | $B^1\Sigma^+\rightarrow A^1\Pi$ (0, 0) | | | $B^1\Sigma^+\rightarrow A^1\Pi$ (0, 3) | | |
|---|---|---|---|---|---|---|
| $J''$ | $P_e\,(J'')$ | $Q_f\,(J'')$ | $R_e\,(J'')$ | $P_e\,(J'')$ | $Q_f\,(J'')$ | $R_e\,(J'')$ |
| 1 | 22159.76 [wb] | 22163.32 [b] | 22170.60 [wb] | 17960.852 | 17964.481 | 17971.728 |
| 2 | 22157.590 | 22164.72 [b] | 22175.717 | 17958.758 | 17966.010 | 17976.886 |
| 3 | 22156.280 | 22166.84 [b] | 22181.662 | 17957.436 | 17968.310 | 17982.811 |
| 4 | 22155.944 | 22169.721 | 22188.573 | 17956.87 [b] | 17971.374 | 17989.498 |
| 5 | 22156.963 | 22173.42 [b] | 22196.853 | 17957.074 | 17975.20 [b] | 17996.969 |
| 6 | 22147.991 | 22178.013 | 22195.16 [b] | 17958.043 | 17979.817 | 18005.218 |
| 7 | 22150.998 | 22183.728 | 22205.38 [b] | 17959.802 | 17985.19 [b] | 18014.18 [b] |
| 8 | 22153.749 | 22190.78 [b] | 22215.342 | 17962.339 | 17991.322 | 18023.931 |
| 9 | 22156.698 | 22182.374 | 22225.560 | 17965.595 | 17998.229 | 18034.457 |
| 10 | 22160.24 [b] | 22190.78 [b] | 22236.344 | 17969.647 | 18005.912 | 18045.75 [b] |
| 11 | 22164.62 [b] | 22199.11 [b] | 22247.964 | 17974.476 | 18014.378 | 18057.816 |
| 12 | 22170.255 | 22207.661 | 22260.829 | 17980.095 | 18023.654 | 18070.668 |
| 13 | 22165.997 | 22216.589 | 22263.797 | 17986.490 | 18033.79 [b] | 18084.293 |
| 14 | 22173.42 [b] | 22226.000 | 22278.472 | 17993.710 | 18045.329 | 18098.740 |
| 15 | 22180.607 | 22235.959 | 22292.854 | 18001.828 | 18054.479 | 18114.079 |
| 16 | 22187.95 [b] | 22246.504 | 22307.416 | 18011.209 | 18067.406 | 18130.677 |
| 17 | 22195.722 | 22257.670 | 22322.404 | 18016.149 | 18080.649 | 18142.829 |
| 18 | 22204.059 | 22269.485 | 22337.954 | 18028.288 | 18094.594 | 18162.178 |
| 19 | 22213.056 | 22282.006 | 22354.144 | 18039.67 [b] | 18109.531 | 18180.770 |
| 20 | 22223.03 [b] | 22295.519 | 22371.317 | 18051.579 | 18122.03 [b] | 18199.867 |
| 21 | 22231.94 [b] | 22307.998 | 22387.43 [b] | 18064.156 | 18139.766 | 18219.646 |
| 22 | 22243.48 [b] | 22323.108 | 22406.144 | 18077.480 | 18156.885 | 18240.151 |
| 23 | 22255.776 | 22338.982 | 22425.622 | 18091.557 | 18174.610 | 18261.402 |
| 24 | 22270.567 | 22357.333 | 22447.60 [b] | 18106.390 | 18193.058 | 18283.405 |
| 25 | 22274.727 | 22365.085 | 22458.93 [b] | 18121.97 [b] | 18212.258 | 18306.173 |
| 26 | 22290.412 | 22384.336 | 22481.76 [b] | 18138.361 | 18232.214 | 18329.702 |
| 27 | 22306.03 [w] | 22403.538 | | 18155.503 | 18252.94 [b] | 18354.001 |
| 28 | | 22425.06 [wb] | | 18173.39 [b] | 18274.436 | 18379.057 |
| 29 | | | | 18192.108 | 18296.699 | 18404.887 |
| 30 | | | | 18211.56 [b] | 18319.73 [b] | 18431.483 |
| 31 | | | | 18231.824 | 18343.537 | 18458.85 [b] |
| 32 | | | | 18252.84 [b] | 18368.127 | 18486.97 [b] |
| 33 | | | | 18274.657 | 18393.488 | 18515.90 [wb] |
| 34 | | | | 18297.291 | 18419.678 | 18545.63 [wb] |
| 35 | | | | 18320.53 [w] | 18446.46 [b] | 18575.95 [wb] |
| 36 | | | | | 18474.16 [b] | 18607.19 [wb] |

[a] All values in $cm^{-1}$.
[b] Lines marked with '*w*' and/or '*b*' were weak and/or blended, respectively.
[c] The estimated absolute calibration $1\sigma$ uncertainty was 0.003 $cm^{-1}$. The expected absolute accuracy of the line frequency measurements was 0.003 $cm^{-1}$ for the medium-strength and single lines as well as 0.01-0.02 $cm^{-1}$ for the weak/blended lines.

## Table 2

($C^1\Sigma^+$, $B^1\Sigma^+$ ) → ($e^3\Sigma^-$, $d^3\Delta_i$, $a'^3\Sigma^+$, $I^1\Sigma^-$, and $D^1\Delta$) transitions arising from perturbations of the $A^1\Pi$ ($\upsilon$ = 0 - 3) levels in $^{13}C^{17}O$.[a, b]

| System | Band | *J''* | Branch / Wavenumber (cm$^{-1}$) | | | | | |
|---|---|---|---|---|---|---|---|---|
| $B^1\Sigma^+ \rightarrow e^3\Sigma^-$ | (1, 1) [c] | 4 | $^qP_{11ee}$ | *24151.28* [wb] | | | | |
| | | 5 | | *24154.85* [wb] | $^qQ_{12ef}$ | *24161.09* [wb] | | *24194.23* [wb] |
| | | 6 | | 24170.326 | | *24167.77* [b] | $^sR_{11ee}$ | *24216.87* [b] |
| | | 7 | | *24175.73* [b] | | 24175.250 | | *24229.44* [b] |
| | | 8 | $^oP_{13ee}$ | 24140.18 [wb] | | 24183.193 | | |
| | | 9 | | 24145.16 [wb] | | 24208.545 | | |
| | | 10 | | 24151.262 | | 24218.995 | | |
| | | 11 | | 24158.27 [b] | | 24231.37 [w] | | |
| | | 12 | | *24165.84* [b] | | | $^qR_{13ee}$ | *24255.25* [b] |
| | | 13 | | *24185.171* | | | | *24281.720* |
| | | 14 | | *24194.66* [b] | | | | *24298.35* [wb] |
| | | 15 | | *24206.33* [b] | | | | |
| | (0, 1) | 2 | $^qP_{11ee}$ | 22135.25 [w] | $^qQ_{12ef}$ | 22136.38 [b] | | |
| | | 3 | | 22137.46 [b] | | 22140.06 [b] | | |
| | | 4 | | 22140.608 | | 22144.87 [b] | $^sR_{11ee}$ | 22173.23 [wb] |
| | | 5 | | 22144.361 | | 22150.82 [b] | | 22184.24 [wb] |
| | | 6 | | | | 22157.78 [b] | | |
| | | 7 | | | | 22165.603 | | |
| | | 8 | $^oP_{13ee}$ | 22130.53 [wb] | | 22173.907 | | |
| | | 9 | | 22135.89 [wb] | | 22199.69 [b] | | |
| | | 10 | | 22142.39 [wb] | | 22210.596 | $^qR_{13ee}$ | 22218.49 [b] |
| | | 11 | | 22149.871 | | 22223.456 | | 22233.203 |
| | | 12 | | 22157.98 [b] | | 22238.06 [b] | | 22248.541 |
| | | 13 | | 22177.86 [b] | | 22254.20 [wb] | | 22275.66 [b] |
| | | 14 | | 22187.95 [b] | | | | 22292.99 [w] |
| | | 15 | | | | | | 22312.52 [wb] |
| | (0, 4) | 27 | | | $^qQ_{12ef}$ | 19590.088 | | |
| | | 28 | | | | 19637.20 [b] | | |
| $B^1\Sigma^+ \rightarrow d^3\Delta$ | (1, 4) | 20 | | | $^rQ_{11ef}$ | 24286.66 [wb] | | |
| | | 21 | | | | | | |
| | | 22 | | | $^qQ_{12ef}$ | 24281.56 [wb] | | |
| | | 23 | $^pP_{12ee}$ | 24228.09 [wb] | | | | |
| | | 24 | | | | | $^qR_{13ee}$ | 24427.23 [wb] |
| | (0, 4) | 21 | | | $^rQ_{11ef}$ | 22316.56 [wb] | | |
| | | 22 | | | | | | |
| | | 23 | | | | | $^qR_{13ee}$ | 22398.85 [b] |
| | | 24 | | | $^qQ_{12ef}$ | 22341.319 | | |
| | | 25 | | | | 22383.13 [wb] | | |
| | | 26 | | | | 22415.60 [wb] | | |
| | (0, 7) | 31 | | | $^rQ_{11ef}$ | 19668.59 [b] | | |
| | | 32 | | | | 19716.43 [b] | | |
| | | 33 | $^qP_{11ee}$ | 19645.78 [b] | | | | |
| | (0, 8) | 10 | $^qP_{11ee}$ | 17991.92 [b] | | | | |
| | | … | | | | | | |
| | | 16 | $^oP_{13ee}$ | 17991.88 [b] | | | | |
| $B^1\Sigma^+ \rightarrow a'^3\Sigma^+$ | (0, 10) | 2 | | | $^pQ_{13ef}$ | 20734.67 [w] | | |
| | | 3 | | | | 20736.98 [w] | | |
| | | 4 | $^pP_{12ee}$ | 20738.249 | | 20740.61 [wb] | | |
| | (0, 13) | 14 | | | $^rQ_{11ef}$ | 18036.857 | | |
| | | 15 | | | | 18064.289 | | |
| | | 16 | $^pP_{12ee}$ | 17997.984 | | | $^rR_{12ee}$ | 18117.449 |
| | | 17 | | 18024.949 | | | | 18151.630 |
| | | 18 | | 18046.91 [b] | | | | |
| | | 19 | | | $^pQ_{13ef}$ | 18096.55 [w] | | |
| | | 20 | | | | 18128.001 | | |
| $C^1\Sigma^+ \rightarrow a'^3\Sigma^+$ | (0, 10) | 1 | | | $^pQ_{13ef}$ | 25735.38 [b] | | |
| | | 2 | | | | 25736.48 [b] | | |
| | | 3 | | | | 25738.77 [b] | | |
| | | 4 | | | | 25742.34 [b] | | |
| | | 13 | | | $^rQ_{11ef}$ | 23015.31 [b] | | |
| | (0, 13) | 14 | | | | 23037.92 [b] | | |
| | | 15 | | | | 23065.26 [b] | $^rR_{12ee}$ | 23090.90 [b] |
| | | 16 | $^pP_{12ee}$ | 22998.92 [wb] | | 23091.11 [b] | | 23118.17 [b] |
| | | 17 | | 23025.82 [wb] | | | | 23152.26 [b] |
| | | 18 | | | | | | |
| | | 19 | | | | | | |
| | | 20 | | | $^{p}Q_{13ef}$ | 23128.36 $^{b}$ | | |
| | | 21 | | | | | | |
| | | 22 | | | | 23187.95 $^{b}$ | | |
| $B^1\Sigma^+ \rightarrow I^1\Sigma^-$ | (0, 2) | 35 | | | $^{q}Q_{11ef}$ | 21148.89 $^{b}$ | | |
| | | 36 | | | | 21199.64 $^{b}$ | | |
| | (0, 3) | 7 | | | $^{q}Q_{11ef}$ | 19347.351 | | |
| | | 8 | | | | 19358.594 | | |
| | (0, 5) | 40 | | | $^{q}Q_{11ef}$ | 18556.25 $^{wb}$ | | |
| | | 41 | | | | 18615.13 $^{wb}$ | | |
| $C^1\Sigma^+ \rightarrow I^1\Sigma^-$ | (0, 3) | 7 | | | $^{q}Q_{11ef}$ | 24348.96 $^{b}$ | | |
| | | 8 | | | | 24360.149 | | |
| $B^1\Sigma^+ \rightarrow D^1\Delta$ | (0, 1) | 25 | | | $^{q}Q_{11ef}$ | 20903.11 $^{b}$ | | |
| | | 26 | | | | 20937.80 $^{w}$ | | |
| | | 27 | | | | 20973.72 $^{wb}$ | $^{r}R_{11ee}$ | 21074.75 $^{wb}$ |
| | | 28 | $^{p}P_{11ee}$ | 20910.76 $^{b}$ | | 21011.86 $^{wb}$ | | 21116.47 $^{w}$ |
| | (0, 4) | 31 | | | $^{q}Q_{11ef}$ | 18271.42 $^{wb}$ | | |
| | | 32 | | | | 18317.20 $^{wb}$ | | |
| | | 33 | | | | | | |
| | | 34 | $^{p}P_{11ee}$ | 18290.62 $^{w}$ | | | $^{r}R_{11ee}$ | 18538.961 |
| $C^1\Sigma^+ \rightarrow D^1\Delta$ | (0, 1) | 27 | $^{p}P_{11ee}$ | 25875.66 $^{b}$ | | | $^{r}R_{11ee}$ | 26073.80 $^{b}$ |
| | | 28 | | 25910.06 $^{b}$ | | | | 26115.35 $^{b}$ |

[a] All values in $cm^{-1}$.
[b] Lines marked with '*w*' and/or '*b*' were weak and/or blended, respectively. The estimated absolute calibration $1\sigma$ uncertainty was 0.003 $cm^{-1}$. The expected absolute accuracy of the line frequency measurements was 0.003 $cm^{-1}$ for the medium-strength and single lines as well as 0.01-0.02 $cm^{-1}$ for the weak/blended lines. The left-superscripts *o*, *p*, *r*, *s*, or *q* denote the change of the total angular momentum excluding spin of the perturber states [88].
[c] The wavenumbers of the $^{13}C^{17}O$ $B^1\Sigma^+ \rightarrow e^3\Sigma^-$ (1, 1) band were already published by us in Ref. [34]. In this table, all the lines of the band under consideration are presented which were used in the current deperturbation fit. Some of the line assignments, whose values are given in italics, were corrected in relation to Ref. [34]. It is worth noticing a relevant change in the interpretation of the $J''$ = 13 lines, that is, lines 24185.1712 $cm^{-1}$ and 24281.7195 $cm^{-1}$ thought in Ref. [34] to belong to the $^{13}C^{17}O$ $B^1\Sigma^+ \rightarrow A^1\Pi$ (1, 0) band, however, turned out to be its extra-lines, that is $^{o}P_{13ee}(13)$ and $^{q}R_{13ee}(13)$ of the $B^1\Sigma^+ \rightarrow e^3\Sigma^-$ (1, 1) transition. Thus, lines 24173.3064 $cm^{-1}$ and 24269.8541 $cm^{-1}$ [34] turned out to be the $P_{11ee}(13)$ and $R_{11ee}(13)$ lines belonging to the $B^1\Sigma^+ \rightarrow A^1\Pi$ (1, 0) band.

**Table 3**.

Extended and corrected assignment [a] of the heavily perturbed and/or extremely weak lines of the $^{13}C^{17}O$ $B^1\Sigma^+ \rightarrow A^1\Pi$ and $C^1\Sigma^+ \rightarrow A^1\Pi$ systems.[b]

| | $B^1\Sigma^+ \rightarrow A^1\Pi$ (0, 1) | | | $B^1\Sigma^+ \rightarrow A^1\Pi$ (1, 0) | | | $B^1\Sigma^+ \rightarrow A^1\Pi$ (1, 1) | | $C^1\Sigma^+ \rightarrow A^1\Pi$ (0, 1) | |
|---|---|---|---|---|---|---|---|---|---|---|
| $J''$ | $P_{11ee}(J'')$ | $Q_{11ef}(J'')$ | $R_{11ee}(J'')$ | $P_{11ee}(J'')$ | $Q_{11ef}(J'')$ | $R_{11ee}(J'')$ | $P_{11ee}(J'')$ | $Q_{11ef}(J'')$ | $P_{11ee}(J'')$ | $R_{11ee}(J'')$ |
| 1 | | | | | | | | | | |
| 2 | | | | 24168.50 $^{wb}$ | | 24186.40 $^{w}$ | | | | |
| 3 | | | | | | | | | | |
| 4 | | | | | | | | | | |
| 5 | | | | | | 24206.85 $^{b}$ | | | | |
| 6 | 20720.011 | | 20767.15 $^{b}$ | | | | | 22751.599 | 25721.72 $^{b}$ | 25768.759 |
| 7 | | | | | 24193.36 $^{b}$ | | | | | |
| 8 | 20723.16 $^{b}$ | | | | | | | | | |
| … | | | | | | | | | | |
| 13 | | | | 24173.31 $^{b}$ | | 24269.85 $^{b}$ | | | | |
| … | | | | | | | | | | |
| 17 | | | | 24200.32 $^{b}$ | | 24325.35 $^{b}$ | | | | |
| 18 | | | | | | | 22783.20 $^{b}$ | | | |
| 19 | | | | | | | 22791.741 | | | |
| 20 | 20798.86b | | 20947.157 | | | | | | 25799.30 $^{w}$ | 25947.32 $^{wb}$ |
| 21 | | | | 24233.02 $^{b}$ | **24308.05** $^{b}$ | | **22810.72** $^{w}$ | | | |
| 22 | | | | **24243.58** $^{wb}$ | | | **22821.17** $^{w}$ | | | |
| 23 | | | | **24254.86** $^{b}$ | | | | | | |
| 24 | | | | **24268.54** $^{b}$ | | | | | | |
| 25 | 20859.569 | | 21043.76 $^{b}$ | **24271.57** $^{b}$ | | | | | | |
| 26 | 20873.795 | | 21065.15 $^{b}$ | | | | | | | |
| 27 | | 20986.328 | | | | | | | | |
| 28 | 20903.69 $^{w}$ | 21004.76 $^{wb}$ | 21109.365 | | | | | | | |
| 29 | 20920.255 | 21024.973 | 21133.09 $^{b}$ | | | | | | | |
| 30 | 20937.300 | 21045.543 | | | | | | | | |
| 31 | 20954.98 $^{b}$ | 21066.813 | | | | | | | | |
| 32 | **20973.34** $^{b}$ | 21088.782 | | | | | | | | |
| 33 | **20992.366** | 21111.478 | | | | | | | | |
| 34 | | 21134.94 $^{b}$ | | | | | | | | |
| 35 | | **21159.96** $^{b}$ | | | | | | | | |

| | $B^1\Sigma^+ \rightarrow A^1\Pi$ (0, 2) | | | $C^1\Sigma^+ \rightarrow A^1\Pi$ (0, 2) | | | $C^1\Sigma^+ \rightarrow A^1\Pi$ (0, 3) | | |
|---|---|---|---|---|---|---|---|---|---|
| $J''$ | $P_{11ee}(J'')$ | $Q_{11ef}(J'')$ | $R_{11ee}(J'')$ | $P_{11ee}(J'')$ | $Q_{11ef}(J'')$ | $R_{11ee}(J'')$ | $P_{11ee}(J'')$ | $Q_{11ef}(J'')$ | $R_{11ee}(J'')$ |
| 1 | | | | | 24332.63 $^{wb}$ | | 22962.67 $^{wb}$ | | 22973.53 $^{wb}$ |
| 2 | | | | | 24334.06 $^{w}$ | | 22960.57 $^{w}$ | | 22978.66 $^{wb}$ |
| 3 | | | | | 24336.22 $^{w}$ | | 22959.23 $^{b}$ | | 22984.57 $^{wb}$ |
| 4 | | | | | 24339.09 $^{w}$ | | 22958.64 $^{b}$ | | 22991.216 |
| 5 | | | | | 24342.678 | | | | |
| 6 | | | | | | | | | |
| 7 | | | | | 24352.08 $^{b}$ | | | | |
| 8 | | | | | | | | | |
| 9 | | 19362.60 $^{b}$ | | | | | | | |
| … | | | | | | | | | |
| 18 | | | | | | | | 23095.216 | |
| 19 | | | | | | | | 23109.99 $^{b}$ | |
| 20 | | | | | | | | 23122.3558 | |
| 21 | | | | | | | | 23139.9881 | |
| 22 | | | | 24434.49 $^{wb}$ | | | | 23156.9625 | |
| 23 | 19447.82 $^{b}$ | | 19617.68 $^{b}$ | 24447.911 | | **24617.46** $^{wb}$ | | **23174.5270** | |
| 24 | | | 19640.782 | 24463.49 $^{wb}$ | | **24640.19** $^{wb}$ | | **23192.8315** | |
| 25 | 19473.191 | | 19657.374 | 24472.82 $^{wb}$ | | | | **23211.89** $^{b}$ | |
| 26 | | | 19680.883 | 24489.11 $^{wb}$ | 24585.37 $^{b}$ | | | **23231.68** $^{b}$ | |
| 27 | 19505.94 $^{b}$ | | 19704.435 | | **24607.04** $^{b}$ | | | **23252.18** $^{b}$ | |
| 28 | 19522.954 | 19619.564 | 19728.597 | | **24618.66** $^{b}$ | | | **23273.52** $^{b}$ | |
| 29 | 19540.750 | 19642.995 | 19753.529 | | | | | | |
| 30 | 19559.624 | 19665.54 $^{b}$ | 19779.526 | | | | | | |
| 31 | 19581.584 | 19688.515 | **19808.61** $^{w}$ | | | | | | |
| 32 | | 19711.44 $^{b}$ | | | | | | | |
| 33 | | 19736.081 | | | | | | | |

[a] An extension (in bold) and improvement over the previously analyzed [33-35] data. All values in $cm^{-1}$.
[b] Lines marked with '*w*' and/or '*b*' were weak and/or blended, respectively. The estimated absolute calibration $1\sigma$ uncertainty was 0.003 $cm^{-1}$. The expected absolute accuracy of the line frequency measurements was 0.003 $cm^{-1}$ for the medium-strength and single lines as well as 0.01-0.02 $cm^{-1}$ for the weak/blended lines.

## Table 4

Deperturbed rovibrational constants of the $A^1\Pi$ ($\upsilon$ = 0, 1, 2, 3) levels as well as their perturbers in $^{13}C^{17}O$.[a]

| Constant / Level | $A^1\Pi(\upsilon = 0)$ | $A^1\Pi(\upsilon = 1)$ | $A^1\Pi(\upsilon = 2)$ | $A^1\Pi(\upsilon = 3)$ |
|---|---|---|---|---|
| $T_\upsilon$ | 64757.7221 (10) | 66188.7843 (34) | 67587.3349 (44) | 68953.3740 (19) |
| $B_\upsilon$ | 1.493 035 (18) | 1.472 586 (16) | 1.451 969 (14) | 1.431 118 (11) |
| $q_\upsilon \times 10^5$ | 1.77 (59) | -1.63 (41) | -1.25 [c] | -1.35 (32) |
| $D_\upsilon \times 10^6$ | 6.523 (28) | 6.399 (17) | 6.480 (21) | 6.558 (14) |
| $H_\upsilon \times 10^{11}$ | -1.10 [b] | -1.10 [b] | -1.10 [b] | -1.10 [b] |

| Constant / Level | $d^3\Delta_i(\upsilon = 4)$ | $d^3\Delta_i(\upsilon = 5)$ | $d^3\Delta_i(\upsilon = 7)$ | $d^3\Delta_i(\upsilon = 8)$ |
|---|---|---|---|---|
| $T_\upsilon$ | 64967.897 (47) | 66009.64 [h] | 68028.906 (84) | 69011.413 (11) |
| $B_\upsilon$ | 1.150 629 (78) | 1.135 07 [f] | 1.104 643 (87) | 1.089 99 [f] |
| $A_\upsilon$ | -15.947 (16) | -16.13 [f] | -16.42 [f] | -16.57 [f] |
| $\lambda_\upsilon$ | 0.772 [d] | 0.865 [d] | 1.052 [d] | 1.145 [d] |
| $\gamma_\upsilon \times 10^3$ | -7.76 [e] | -7.76 [e] | -7.76 [e] | -7.76 [e] |
| $D_\upsilon \times 10^6$ | 5.61 [f] | 5.59 [f] | 5.57 [f] | 5.56 [f] |
| $H_\upsilon \times 10^{13}$ | -6.45 [g] | -6.45 [g] | -6.45 [g] | -6.45 [g] |
| $A_{D\upsilon} \times 10^5$ | -4.72 [e] | -4.72 [e] | -4.72 [e] | -4.72 [e] |
| $\eta$ | -21.598 (12) | -15.98 [i] | 9.57 (23) | -0.560 (59) |

| Constant / Level | $e^3\Sigma^-(\upsilon = 1)$ | $e^3\Sigma^-(\upsilon = 3)$ | $e^3\Sigma^-(\upsilon = 4)$ | $e^3\Sigma^-(\upsilon = 5)$ |
|---|---|---|---|---|
| $T_\upsilon$ | 64781.49523 (86) | 66842.0375 [j] | 67843.866 (65) | 68829.515 [j] |
| $B_\upsilon$ | 1.170 759 (17) | 1.139 058 [f] | 1.123 537 (65) | 1.108 336 [f] |
| $\lambda_\upsilon$ | 0.5139 (35) | 0.528 [d] | 0.626 (32) | 0.561 [d] |
| $D_\upsilon \times 10^6$ | 5.90 [f] | 5.83 [f] | 5.79 [f] | 5.77 [f] |
| $H_\upsilon \times 10^{12}$ | -1.61 [g] | -1.61 [g] | -1.61 [g] | -1.61 [g] |
| $\eta$ | 14.5737 (28) | 4.748 [i] | 13.101 (12) | 8.322 [i] |

| Constant / Level | $a'^3\Sigma^+(\upsilon = 9)$ | $a'^3\Sigma^+(\upsilon = 10)$ | $a'^3\Sigma^+(\upsilon = 12)$ | $a'^3\Sigma^+(\upsilon = 13)$ |
|---|---|---|---|---|
| $T_\upsilon$ | 65182.01 [j] | 66180.0621 (83) | 68123.35 [j] | 69068.9897 (45) |
| $B_\upsilon$ | 1.104 62 [f] | 1.089 35 (47) | 1.061027 [f] | 1.046 459 (21) |
| $\lambda_\upsilon$ | -1.115 [d] | -1.0955 (78) | -1.096 [d] | -1.111 8 (67) |
| $\gamma_\upsilon \times 10^3$ | -5.98 [e] | -5.98 [e] | -5.98 [e] | -5.98 [e] |
| $D_\upsilon \times 10^6$ | 5.43 [f] | 5.43 [f] | 5.41 [f] | 5.41 [f] |
| $H_\upsilon \times 10^{13}$ | -3.22 [g] | -3.22 [g] | -3.22 [g] | -3.22 [g] |
| $\eta$ | -2.60 [i] | -5.157 (15) | 6.02 (22) | 7.2554 (59) |

| Constant / Level | $I^1\Sigma^-(\upsilon = 0)$ | $I^1\Sigma^-(\upsilon = 2)$ | $I^1\Sigma^-(\upsilon = 3)$ | $I^1\Sigma^-(\upsilon = 5)$ |
|---|---|---|---|---|
| $T_\upsilon$ | 64565.40 [k] | 66613.073 (18) | 67608.1757 (98) | 69541.194 (31) |
| $B_\upsilon$ | 1.173 59 [f] | 1.141 89 [f] | 1.126 04 (18) | 1.093 99 [f] |
| $D_\upsilon \times 10^6$ | 5.93 [g] | 5.97 [g] | 5.98 [g] | 6.02 [g] |
| $H_\upsilon \times 10^{12}$ | 2.42 [g] | 2.42 [g] | 2.42 [g] | 2.42 [g] |
| $\xi \times 10^2$ | -3.92 [i] | -7.492 (32) | -5.264 (61) | 2.52 (16) |

| Constant / Level | | $D^1\Delta(\upsilon = 1)$ | | $D^1\Delta(\upsilon = 4)$ |
|---|---|---|---|---|
| $T_\upsilon$ | | 66447.8209 (50) | | 69355.14 (11) |
| $B_\upsilon$ | | 1.145 99 [l] | | 1.098 394 (97) |
| $D_\upsilon \times 10^6$ | | 6.08 [l] | | 6.03 [l] |
| $H_\upsilon \times 10^{13}$ | | -2.42 [g] | | -2.42 [g] |
| $\xi \times 10^2$ | | -6.295 (38) | | -2.21 (11) |

[a] All values in $cm^{-1}$. Numbers in parentheses denote one standard deviation in units of the last significant digit. $T_\upsilon$ denotes ($\upsilon$ = 0, $J$ = 0) term value of the excited state in relation to the $X^1\Sigma^+$ ($\upsilon$ = 0, $J$ = 0) ground state; $\eta_{\upsilon,\upsilon}$-spin-orbit interaction parameter; $\xi_{\upsilon,\upsilon}$-$L$-uncoupling interaction parameter. The values without given uncertainties were fixed during the final fit.

[b] Isotopically recalculated on the basis of values given by Le Floch [45].

[c] Isotopically recalculated on the basis of the individual constants $q$ of $A^1\Pi$ ($\upsilon$ = 0-4) by Niu et al.[47,48].

[d] Isotopically recalculated on the basis of the spin-spin $C$ constants of Field [42] taking into account the equation $\lambda = -\left(\frac{3}{2}\right)C$ (see Table 3.4 in Ref. [36]).

[e] Calculated on the basis of data by Field [42] using the 29979,2458 MHz/$cm^{-1}$ conversion [89] and isotopically recalculated.

[f] Isotopically recalculated on the basis of values given by Field [42].

[g] Isotopically recalculated on the basis of values given by Le Floch [44].

[h] Evaluated by the means of the isotopically recalculated vibrational equilibrium constants of $d^3\Delta_i$ by Field [42], and $X^1\Sigma^+$ by Le Floch [89]. $T_e$ of $d^3\Delta_i$ was taken from Huber and Herzberg [76].

[i] Calculated on the basis of the isotopologue-independent electronic factors $\boldsymbol{a}$ and $\boldsymbol{b}$ given by Field [43,90] and Le Floch [44] as well as $\langle \upsilon_A | \upsilon_{d,e,\ \text{or}\ a'} \rangle$ and $\langle \upsilon_A | B(r) | \upsilon_{I\ \text{or}\ D} \rangle$ vibrational integrals in $^{13}C^{17}O$, using a method given by Hakalla et al. [55] (eqns. 1-8 in Supplementary Material).

[j] Obtained by means of the isotopically recalculated vibrational equilibrium constants of $e^3\Sigma^-$ or $a'^3\Sigma^+$ by Field [42] and $X^1\Sigma^+$ by Le Floch [89]. $T_e$ of $e^3\Sigma^-$ or $a'^3\Sigma^+$ was taken from Tilford et al. [75].

[k] Calculated on the basis of isotopically recalculated vibrational equilibrium constants of $I^1\Sigma^-$ by Field [42] and $X^1\Sigma^+$ by Le Floch [89]. $T_e$ of $I^1\Sigma^-$ was taken from Herzberg et al. [91].

[l] Isotopically recalculated from Kittrell et al. [74].

**Table 5**

Parameters for the $B^1\Sigma^+$ and $C^1\Sigma^+$ states of $^{13}C^{17}O$.[a]

| Level / Constant | $B^1\Sigma^+(\upsilon = 0)$ | $B^1\Sigma^+(\upsilon = 1)$ | $C^1\Sigma^+(\upsilon = 0)$ |
|---|---|---|---|
| $T_\upsilon$ | 86917.1236 (14) [b] | 88928.0904 (18) [b] | 91918.9416 (17) [b] |
| $B_\upsilon$ | 1.813 234 9 (93)<br>1.813 194 1 (58) [c] | 1.789 919 (21)<br>1.790 227 (23) [d] | 1.809 582 1 (82)<br>1.809 711 3 (93) [e] |
| $D_\upsilon \times 10^6$ | 5.706 (13)<br>5.562 0 (46) [c] | 6.106 (46)<br>6.233 (47) [d] | 5.350 5 (71)<br>5.532 (12) [e] |

[a] All values in $cm^{-1}$. Numbers in parentheses denote one standard deviation in units of the last significant digit.
[b] $T_\upsilon$ denotes position of the level with respect to the $X^1\Sigma^+$ ($\upsilon = 0$, $J = 0$) ground state of $^{13}C^{17}O$. Note that the $T_\upsilon$ uncertainties refer to the fitting with respect to a relative scale (fitting), while the absolute accuracy is not better than the 0.04 $cm^{-1}$.
[c] After Hakalla et al. [33].
[d] After Hakalla et al. [34].
[e] After Hakalla [35].

**Table 6**

Spin-orbit and rotation-electronic perturbation parameters resulting from the deperturbation analysis of the $A^1\Pi$ ($v$ = 0, 1, 2, 3) levels in $^{13}C^{17}O$.[a]

| Interaction | r-centroid [b] (Å) | $\langle v\|v'\rangle$ [b] (unitless) | $\eta_{v,v'}$ [c] (cm$^{-1}$) | $\eta_{v,v'}/\langle v\|v'\rangle$ (cm$^{-1}$) | $\boldsymbol{a}$ [d] (cm$^{-1}$) | $\bar{\boldsymbol{a}}$ [e] (cm$^{-1}$) |
|---|---|---|---|---|---|---|
| $A^1\Pi$ ($v$ = 0) ~ $d^3\Delta_i$ ($v$ = 4) | 1.2288 | 0.3728 | -21.598 (12) | -57.929 (32) | 94.60 (52) | |
| $A^1\Pi$ ($v$ = 2) ~ $d^3\Delta_i$ ($v$ = 7) | 1.2311 | -0.1666 | 9.57 (23) | -57.4 (14) | 93.8 (22) | |
| $A^1\Pi$ ($v$ = 3) ~ $d^3\Delta_i$ ($v$ = 8) | 1.1324 | 0.0107 | -0.560 (59) | -52.2 (56) | 85.3 (90) | |
| | | | | | | **94.60 (52)** [f] |
| | | | | | | 95.59 (27) [g] |
| | | | | | | 96.029 (59) [h] |
| | | | | | | 95.8 [i] |
| $A^1\Pi$ ($v$ = 0) ~ $e^3\Sigma^-$ ($v$ = 1) | 1.3463 | -0.3382 | 14.5737 (28) | -43.0888 (82) | 99.51 (19) | |
| $A^1\Pi$ ($v$ = 2) ~ $e^3\Sigma^-$ ($v$ = 4) | 1.2888 | -0.2998 | 13.101 (12) | -43.700 (41) | 100.92 (95) | |
| | | | | | | **99.56 (19)** [f] |
| | | | | | | 98.90 (33) [g] |
| | | | | | | 99.045 (36) [h] |
| | | | | | | 99.0 [i] |
| | | | | | | 98(4) [j] |
| $A^1\Pi$ ($v$ = 1) ~ $a'^3\Sigma^+$ ($v$ = 10) | 1.1401 | -0.1409 | -5.157 (15) | 36.59 (11) | 84.50 (25) | |
| $A^1\Pi$ ($v$ = 2) ~ $a'^3\Sigma^+$ ($v$ = 12) | 1.1323 | 0.1601 | 6.02 (22) | 37.6 (14) | 86.88 (32) | |
| $A^1\Pi$ ($v$ = 3) ~ $a'^3\Sigma^+$ ($v$ = 13) | 1.1385 | 0.1995 | 7.2554 (59) | 36.364 (30) | 83.98 (68) | |
| | | | | | | **84.02 (66)** [f] |
| | | | | | | 83.62 (12) [g] |
| | | | | | | 82.0 (25) [h] |
| | | | | | | 83.4 [i] |
| | | | | | | 88.6(3) [j] |

| | r-centroid [b] (Å) | $\langle v\|\boldsymbol{B}\|v'\rangle$ [b] (cm$^{-1}$) | $\xi_{v,v'} \times 10^2$ [c] (cm$^{-1}$) | $\xi_{v,v'}/\langle v\|\boldsymbol{B}\|v'\rangle$ (unitless) | $\boldsymbol{b}$ [d] (unitless) | $\bar{\boldsymbol{b}}$ [e] (unitless) |
|---|---|---|---|---|---|---|
| $A^1\Pi$ ($v$ = 1) ~ $I^1\Sigma^-$ ($v$ = 2) | 1.3041 | 0.4526 | -7.492 (32) | -0.1655 (71) | 0.2341 (10) | |
| $A^1\Pi$ ($v$ = 2) ~ $I^1\Sigma^-$ ($v$ = 3) | 1.3178 | 0.3103 | -5.264 (61) | -0.1696 (20) | 0.2399 (28) | |
| $A^1\Pi$ ($v$ = 3) ~ $I^1\Sigma^-$ ($v$ = 5) | 1.3120 | -0.1729 | 2.52 (16) | -0.1458 (93) | 0.206 (13) | |
| | | | | | | **0.2348 (95)** [f] |
| | | | | | | 0.2274 (46) [g] |
| | | | | | | 0.2269 (23) [h] |
| | | | | | | 0.228 [i] |
| | | | | | | 0.212(16) [j] |
| $A^1\Pi$ ($v$ = 1) ~ $D^1\Delta$ ($v$ = 1) | 1.3290 | -0.5531 | -6.294 (38) | 0.1138 (69) | 0.1138 (69) | |
| $A^1\Pi$ ($v$ = 3) ~ $D^1\Delta$ ($v$ = 4) | 1.3141 | -0.1699 | -2.21 (11) | 0.130 (62) | 0.130 (62) | |
| | | | | | | **0.1139 (68)** [f] |
| | | | | | | 0.1103 (14) [g] |
| | | | | | | 0.0969 (44) [h] |
| | | | | | | 0.137(15) [j] |

[a] Numbers in parentheses denote one standard deviation in units of the last significant digit.
[b] The *r*-centroids and vibrational integrals were calculated on the basis of RKRs of the $^{13}C^{17}O$ A, d, e, a', I, and D states obtained from isotopically recalculated equilibrium constants of Field [42], Le Floch et al. [44], Field et al. [36,43], and Kittrell et al. [74]. In order to perform calculations, the program 'FRACONB' of Jung [83] and Jakubek [84] was used.
[c] The spin-orbit and *L*-uncoupling interaction parameters $\eta_{v,v'}$ and $\xi_{v,v'}$ were taken from the global deperturbation fit of $^{13}C^{17}O$ conducted in this work.
[d] The isotopologue-independent spin-orbit and rotation-electronic perturbation parameters $\boldsymbol{a}$ and $\boldsymbol{b}$ were calculated using eqns. 1-8 (Supplementary Material).
[e] The weighted average values of the electronic perturbation parameters.
[f] The present results.
[g] After Hakalla et al. [55] from $^{12}C^{17}O$.
[h] Obtained on the basis of the $^{12}C^{16}O$ data by Le Floch et al. [44].
[i] After Field et al. [43] from $^{12}C^{16}O$.
[j] After Haridass et al. [52] from $^{12}C^{18}O$.

**Table 7**

Term values (in $cm^{-1}$) of the $A^1\Pi$, $\upsilon$ = 0, 1, 2, and 3 rovibronic levels in $^{13}C^{17}O$.[a, b]

| | $A^1\Pi$ ($\upsilon$ = 0) | | $A^1\Pi$ ($\upsilon$ = 1) | | $A^1\Pi$ ($\upsilon$ = 2) | | $A^1\Pi$ ($\upsilon$ = 3) | |
|---|---|---|---|---|---|---|---|---|
| *J* | *e* | *f* | *e* | *f* | *e* | *f* | *e* | *f* |
| 1 | 64757.411 | 64757.393 | 66193.082 | 66193.131 | 67589.927 | 67589.898 | 68956.287 | 68956.234 |
| 2 | 64763.153 | 64763.288 | 66198.866 | 66199.075 | 67595.717 | 67595.750 | 68961.984 | 68962.005 |
| 3 | 64771.710 | 64772.028 | 66207.608 | 66207.882 | 67604.430 | 67604.445 | 68970.568 | 68970.574 |
| 4 | 64782.931 | 64783.652 | 66219.274 | 66219.536 | 67616.056 | 67616.044 | 68982.021 | 68982.000 |
| 5 | 64796.398 | 64798.088 | 66233.891 | 66234.122 | 67630.551 | 67630.567 | 68996.291 | 68996.317 |
| 6 | 64823.529 | 64815.244 | 66251.503 | 66251.644 | 67648.002 | 67647.936 | 69013.468 | 69013.435 |
| 7 | 64842.257 | 64834.947 | 66272.019 | 66272.133 | 67668.297 | 67668.189 | 69033.461 | 69033.485 |
| 8 | 64864.915 | 64856.834 | 66295.498 | 66295.599 | 67691.497 | 67691.673 | 69056.332 | 69056.328 |
| 9 | 64890.945 | 64897.883 | 66321.932 | 66321.987 | 67717.607 | 67717.636 | 69082.050 | 69082.024 |
| 10 | 64920.018 | 64925.755 | 66351.310 | 66351.362 | 67746.609 | 67746.655 | 69110.612 | 69110.591 |
| 11 | 64951.892 | 64957.263 | 66383.646 | 66383.672 | 67778.512 | 67778.534 | 69142.030 | 69141.990 |
| 12 | 64986.119 | 64992.198 | 66418.900 | 66418.933 | 67813.302 | 67813.331 | 69176.281 | 69176.199 |
| 13 | 65033.861 | 65030.362 | 66457.106 | 66457.129 | 67850.988 | 67851.031 | 69213.377 | 69213.163 |
| 14 | 65073.495 | 65071.657 | 66498.232 | 66498.255 | 67891.552 | 67891.603 | 69253.243 | 69252.346 |
| 15 | 65117.055 | 65116.008 | 66542.310 | 66542.305 | 67935.026 | 67935.046 | 69295.846 | 69297.501 |
| 16 | 65164.021 | 65163.412 | 66589.286 | 66589.299 | 67981.338 | 67981.400 | 69340.764 | 69342.528 |
| 17 | 65214.190 | 65213.769 | 66639.201 | 66639.205 | 68030.530 | 68030.580 | 69393.770 | 69390.797 |
| 18 | 65267.370 | 65267.082 | 66692.032 | 66692.019 | 68082.621 | 68082.655 | 69443.164 | 69441.973 |
| 19 | 65323.512 | 65323.311 | 66747.769 | 66747.811 | 68137.538 | 68137.632 | 69496.907 | 69495.802 |
| 20 | 65382.323 | 65382.137 | 66806.512 | 66806.455 | 68195.348 | 68195.436 | 69553.776 | 69555.662 |
| 21 | 65445.721 | 65445.665 | 66868.065 | 66868.079 | 68255.954 | 68256.132 | 69613.521 | 69613.850 |
| 22 | 65510.169 | 65510.047 | 66932.577 | 66932.541 | 68319.351 | 68319.629 | 69676.166 | 69676.250 |
| 23 | 65577.379 | 65577.308 | 66999.967 | 66999.940 | 68385.299 | 68385.973 | 69741.611 | 69741.658 |
| 24 | 65645.723 | 65645.68 $^c$ | 67070.277 | 67070.310 | 68452.609 | 68455.154 | 69809.913 | 69809.935 |
| 25 | 65728.37 $^c$ | 65728.23 $^c$ | 67143.502 | 67143.493 | 68529.803 | 68527.005 | 69881.03 | 69881.018 |
| 26 | 65802.91 $^c$ | 65802.87 $^c$ | 67219.579 | 67219.56 $^c$ | 68603.790 | 68601.259 | 69954.97 $^c$ | 69954.953 |
| 27 | 65881.17 $^c$ | 65881.13 $^c$ | 67298.41 $^c$ | 67298.36 $^c$ | 68681.31 $^c$ | 68676.937 | 70031.70 $^c$ | 70031.796 |
| 28 | | 65960.65 $^c$ | 67380.99 $^c$ | 67380.91 $^c$ | 68761.74 $^c$ | 68766.16 $^c$ | 70111.25 $^c$ | 70111.26 $^c$ |
| 29 | | | 67465.43 $^c$ | 67465.38 $^c$ | 68844.99 $^c$ | 68847.36 $^c$ | 70193.60 $^c$ | 70193.61 $^c$ |
| 30 | | | 67553.05 $^c$ | 67552.99 $^c$ | 68930.73 $^c$ | 68932.99 $^c$ | 70278.74 $^c$ | 70278.75 $^c$ |
| 31 | | | 67643.55 $^c$ | 67643.46 $^c$ | 69016.95 $^c$ | 69021.77 $^c$ | 70366.66 $^c$ | 70366.68 $^c$ |
| 32 | | | 67736.93 $^c$ | 67736.80 $^c$ | 69115.24 $^c$ | 69114.15 $^c$ | 70457.39 $^c$ | 70457.39 $^c$ |
| 33 | | | 67833.21 $^c$ | 67832.96 $^c$ | 69207.89 $^c$ | 69208.37 $^c$ | 70550.86 $^c$ | 70550.87 $^c$ |
| 34 | | | | 67931.91 $^c$ | | 69313.44 $^c$ | 70647.07 $^c$ | 70647.08 $^c$ |
| 35 | | | | 68032.84 $^c$ | | 69403.36 $^c$ | 70746.23 $^c$ | 70746.23 $^c$ |
| 36 | | | | | | | 70847.99 $^c$ | 70848.02 $^c$ |

[a] The rotational terms were calculated with regard to the $X^1\Sigma^+$ ($\upsilon$ = 0, $J$ = 0) ground state from the combined data sets of two experiments: the VIS- and VUV-FT spectroscopy studies of the $^{13}C^{17}O$ B → A, C → A bands (this work and Ref. [33-35]), and B ← X, C ← X transitions (this work and Ref. [58]), respectively. The final term values were obtained using the weighted average method.

[b] Note that the the absolute level energies are only good to 0.04 $cm^{-1}$, but the internal consistency (precision) of this list is much better-for some levels even 0.005 $cm^{-1}$.

[c] Higher absolute term values obtained by means of the $A^1\Pi(\upsilon)$ relative terms (calculated from B → A (0 - $\upsilon''$) [33,34] and C → A (0-$\upsilon''$) [35] bands by means of the Curl and Dane [85] and Watson [86] least-squares method) and fitted $T_\upsilon$ values of the $A^1\Pi(\upsilon)$ levels (from Table 4). The final values are obtained using the weighted average method.

**Table 8**

Term values of the $e^3\Sigma^-$, $a'^3\Sigma^+$, $d^3\Delta$, $I^1\Sigma^-$, and $D^1\Delta$ perturber states of the $A^1\Pi$ ($\upsilon$ = 0, 1, 2, 3) levels in $^{13}C^{17}O$.[a, b]

| | $e^3\Sigma^-$ | | | $a'^3\Sigma^+$ | | | $I^1\Sigma^-$ |
|---|---|---|---|---|---|---|---|
| *J* | $F_{1e}$ | $F_{2f}$ | $F_{3e}$ | $F_{1f}$ | $F_{2e}$ | $F_{3f}$ | $F_{1f}$ |
| | $\upsilon$ = 1 | | | $\upsilon$ = 10 | | | |
| 1 | | | | | | 66187.17 | |
| 2 | 64785.47 | 64791.64 | | | | 66193.33 | |
| 3 | 64790.56 | 64798.82 | | | | 66201.89 | |
| 4 | 64798.28 | 64808.50 | | | 66200.63 | 66212.77 | |
| 5 | 64809.01 | 64820.69 | | | | | |
| 6 | 64811.44 | 64835.47 | | | | | $\upsilon$ = 3 |
| 7 | 64827.49 | 64853.06 | | | | | 67671.33 |
| 8 | | 64873.73 | 64888.13 | | | | 67689.06 |
| 9 | | 64880.57 | 64911.76 | | | | |
| 10 | | 64905.91 | 64937.86 | | | | |
| 11 | | 64932.91 | 64966.65 | | | | |
| 12 | | 64961.80 | 64998.42 | $\upsilon$ = 13 | | | |
| 13 | | 64992.76 | 65021.99 | 69232.80 | | | |
| 14 | | | 65058.98 | 69260.81 | | | |
| 15 | | | 65097.40 | 69287.69 | 69319.89 | | |
| 16 | | | | 69319.68 | 69353.99 | | |
| 17 | | | | | 69384.96 | | |
| 18 | | | | | 69424.53 | | |
| 19 | | | | | | 69508.77 | |
| 20 | | | | | | 69549.66 | |
| 21 | | | | | | | |
| 22 | | | | | | 69645.26 | |
| | $\upsilon$ = 4 | | | | | | |
| 27 | | 68694.59 | | | | | |
| 28 | | 68748.51 | | | | | |
| … | | | | | | | $\upsilon$ = 2 |
| 35 | | | | | | | 68043.82 |
| 36 | | | | | | | 68122.54 |

| | $d^3\Delta$ | | | | | | $D^1\Delta$ | |
|---|---|---|---|---|---|---|---|---|
| | $F_{1e}$ | $F_{1f}$ | $F_{2e}$ | $F_{2f}$ | $F_{3e}$ | $F_{3f}$ | $F_{1e}$ | $F_{1f}$ |
| | $\upsilon$ = 4 | | | | | | $\upsilon$ = 1 | |
| 20 | | 65392.11 | | | | | | |
| 21 | | 65437.10 | | | | | | |
| 22 | | | | 65550.68 | | | | |
| 23 | | | 65604.16 | | | | | |
| 24 | | | | 65661.70 | | | | |
| 25 | | | | 65710.21 | | | | 67190.23 |
| 26 | | | | 65771.64 | | | | 67249.44 |
| 27 | | | | | | | 67310.97 | 67310.96 |
| 28 | $\upsilon$ = 7 | | | | | | 67373.90 | 67373.85 |
| 29 | | | | | | | | |
| 30 | | | | | | | $\upsilon$ = 4 | |
| 31 | | 69041.66 | | | | | | 70438.837 |
| 32 | | 69109.11 | | | | | | 70508.34 |
| 33 | 69179.76 | | | | | | | |
| 34 | | | | | | | 70653.75 | |
| | $\upsilon$ = 8 | | | | | | | |
| 10 | 69088.34 | | | | | | | |
| … | | | | | | | | |
| 16 | | | | | 69360.09 | | | |

[a] All values in $cm^{-1}$.
[b] The rotational terms were calculated with regard to the $X^1\Sigma^+$ ($\upsilon$ = 0, $J$ = 0) ground state from the combined data sets of two experiments: the VIS-FTS and VUV-FTS study of the $^{13}C^{17}O$ B → A, C → A bands (this work and Ref. [33-35]), and B ← X, C ← X transitions (this work and Ref. [58]) with their extra-lines, respectively. The final values of the terms were merged using the weighted average method.

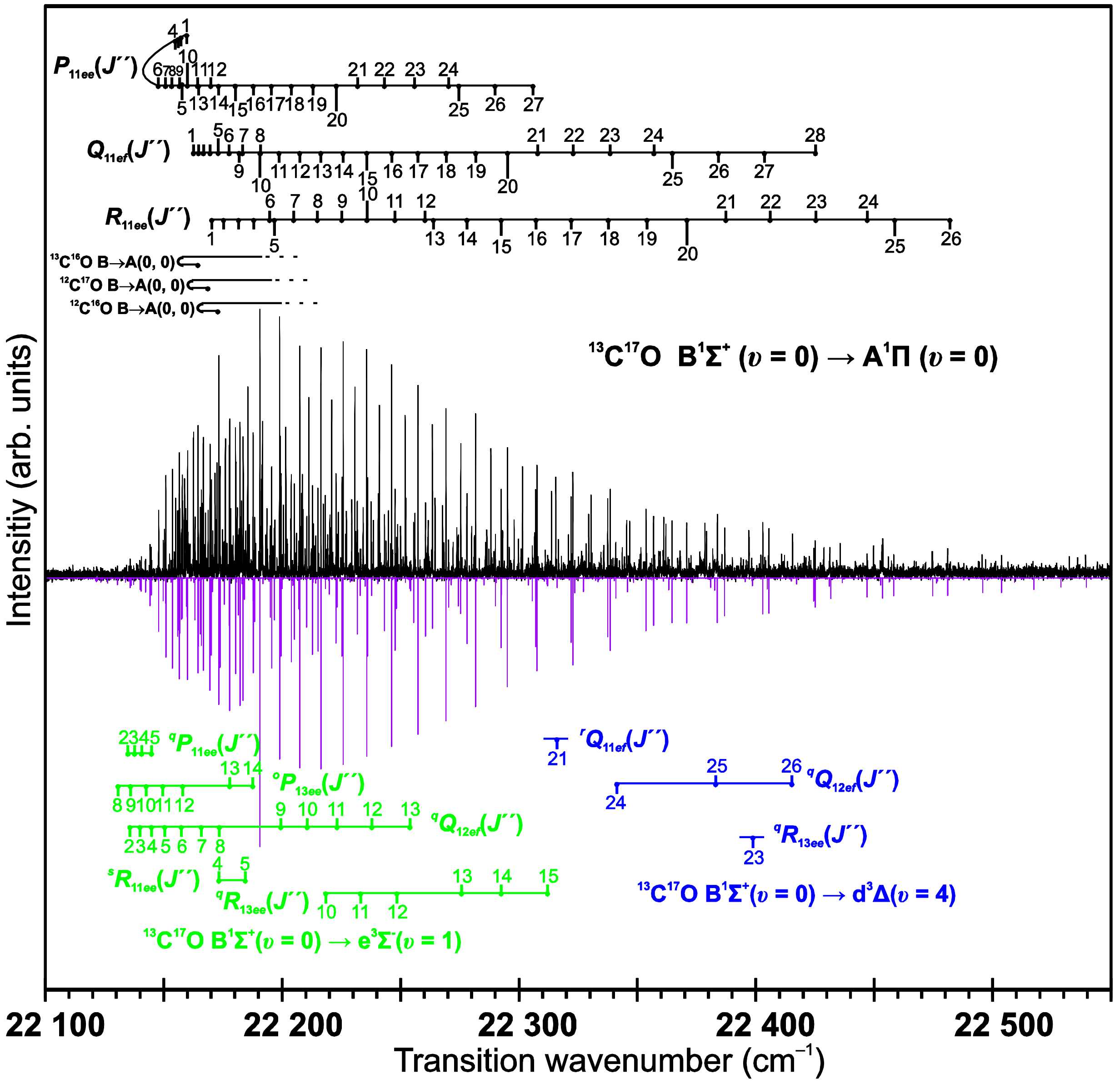

**Fig. 1.** High-resolution emission spectrum (upper trace) of the (0, 0) band of the Ångström ($B^1\Sigma^+ \rightarrow A^1\Pi$) system in the less-abundant $^{13}C^{17}O$ isotopologue recorded in the visible region with the Fourier-transform spectrometer at an instrumental resolution of 0.018 $cm^{-1}$. The extra-lines belong to the nominally forbidden $B^1\Sigma^+ \rightarrow e^3\Sigma^-$ (0, 1), and $B^1\Sigma^+ \rightarrow d^3\Delta_i$ (0, 4) intercombination systems, that are a result of the $A^1\Pi$ ($v$ = 0) ~ $e^3\Sigma^-$ ($v$ = 1), and $A^1\Pi$ ($v$ = 0) ~ $d^3\Delta_i$ ($v$ = 4) interactions, respectively. The lower trace presents a simulated spectrum [79] of the $^{13}C^{17}O$ B - A (0, 0) band. The ratio of the molecular gas composition used to obtain the spectrum was $^{13}C^{17}O$ : $^{13}C^{16}O$ : $^{12}C^{17}O$ : $^{12}C^{16}O$ = 1 : 0.76 : 0.34 : 0.25. The estimated absolute calibration 1$\sigma$ uncertainty was 0.003 $cm^{-1}$. The absolute accuracy of the line frequency measurements was about 0.003 $cm^{-1}$ for the single, medium-strength lines and about 0.02 $cm^{-1}$ for the weakest and/or most blended lines.

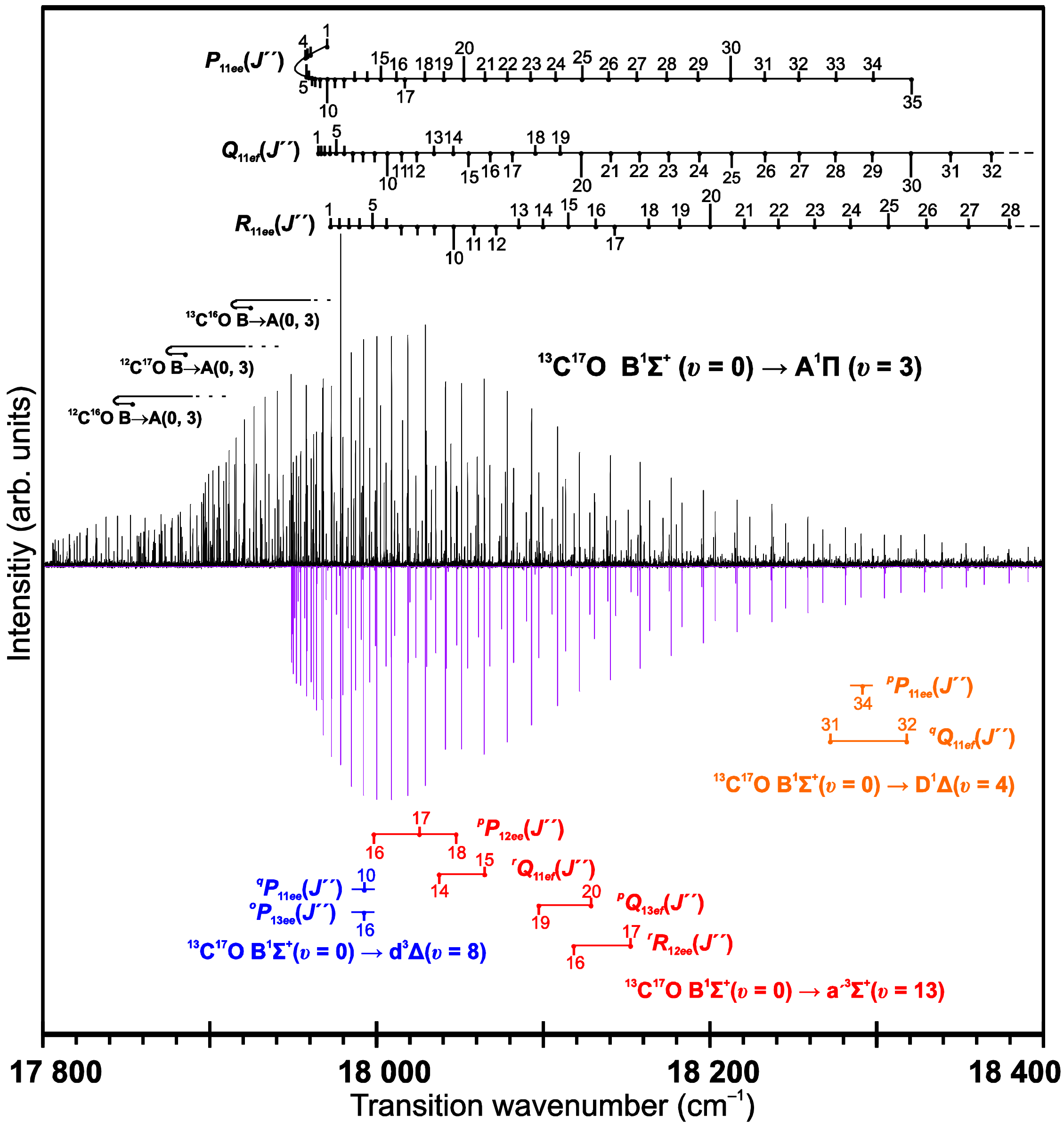

**Fig. 2.** High-resolution emission spectrum (upper trace) of the (0, 3) band of the Ångström ($B^1\Sigma^+ \rightarrow A^1\Pi$) system in the less-abundant $^{13}C^{17}O$ isotopologue recorded in the visible region with the Fourier-transform spectrometer at an instrumental resolution of 0.018 cm$^{-1}$. The extra-lines belong to the $B^1\Sigma^+ \rightarrow D^1\Delta$ (0, 4) system as well as nominally forbidden $B^1\Sigma^+ \rightarrow d^3\Delta_i$ (0, 8) and $B^1\Sigma^+ \rightarrow a'^3\Sigma^+$ (0, 13) intercombination systems, that are a result of the $A^1\Pi$ ($v$ = 3) ~ $D^1\Delta$ ($v$ = 4) as well as $A^1\Pi$ ($v$ = 3) ~ $d^3\Delta_i$ ($v$ = 8), and $A^1\Pi$ ($v$ = 3) ~ $a'^3\Sigma^+$ ($v$ = 13) interactions, respectively. Other details same as in Fig. 1.

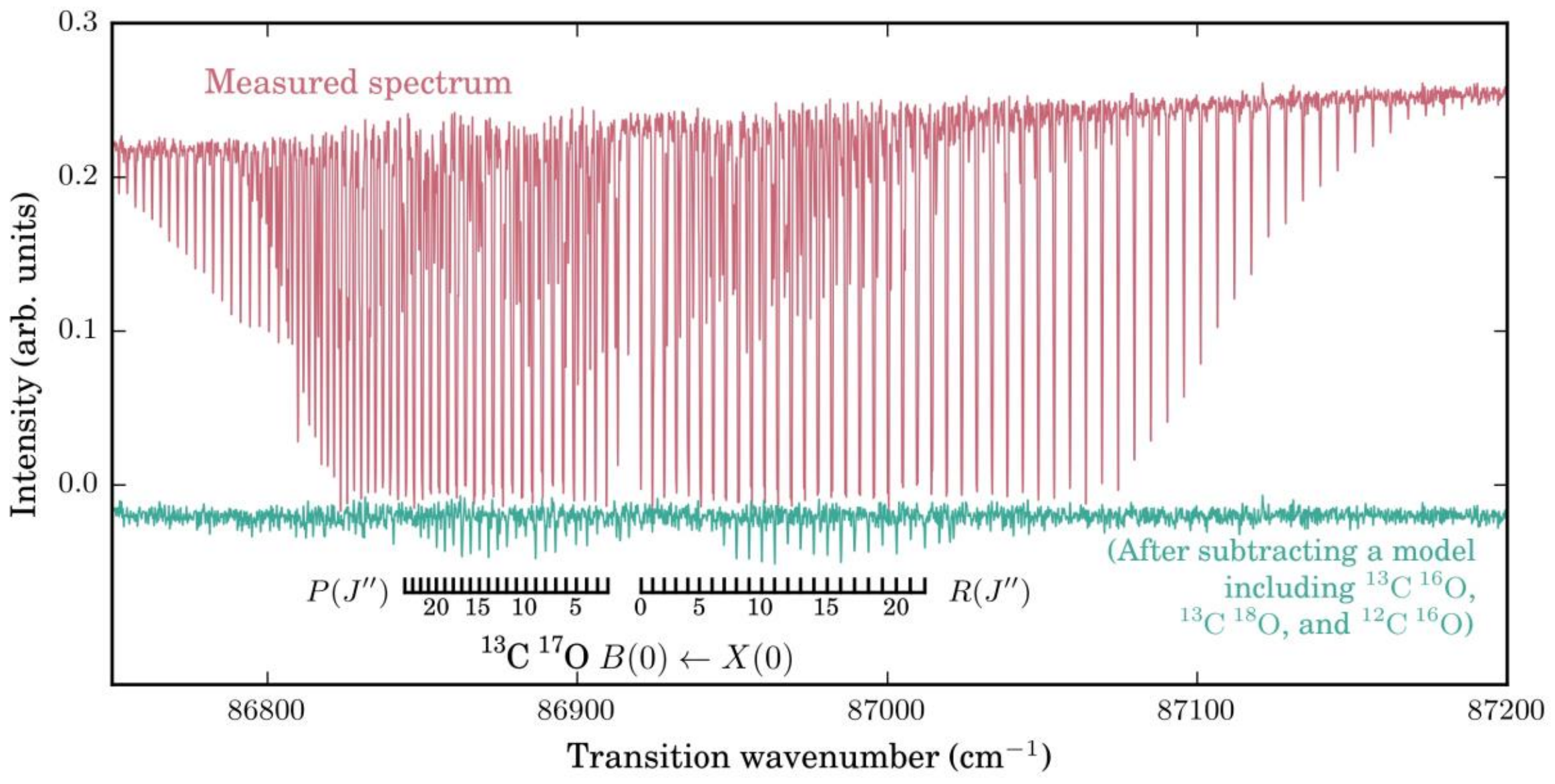

**Fig. 3.** Modelling of the $^{13}C^{17}O$ absorption spectrum (lower trace) of the ($B^1\Sigma^+ \rightarrow X^1\Sigma^+$) (0, 0) band of the Hopfield-Birge system superimposed on the high-resolution absorption spectrum (upper trace) present as impurity in a $^{13}C^{16}O$ gas sample recorded with the SOLEIL VUV-FTS at an instrumental resolution of 0.20 $cm^{-1}$. The ratio of the gases used in the experiment, sorted by concentrations, was $^{13}C^{16}O$ : $^{13}C^{18}O$ : $^{13}C^{17}O$ : $^{12}C^{16}O$ = 1 : 0.041 : 0.073 : 0.0045. The estimated absolute calibration 1σ uncertainty was 0.01 $cm^{-1}$. The statistical uncertainty of fitted line centers was about 0.003 $cm^{-1}$ for the unblended, medium-strength lines and about 0.09 $cm^{-1}$ for the weakest and/or most blended lines.

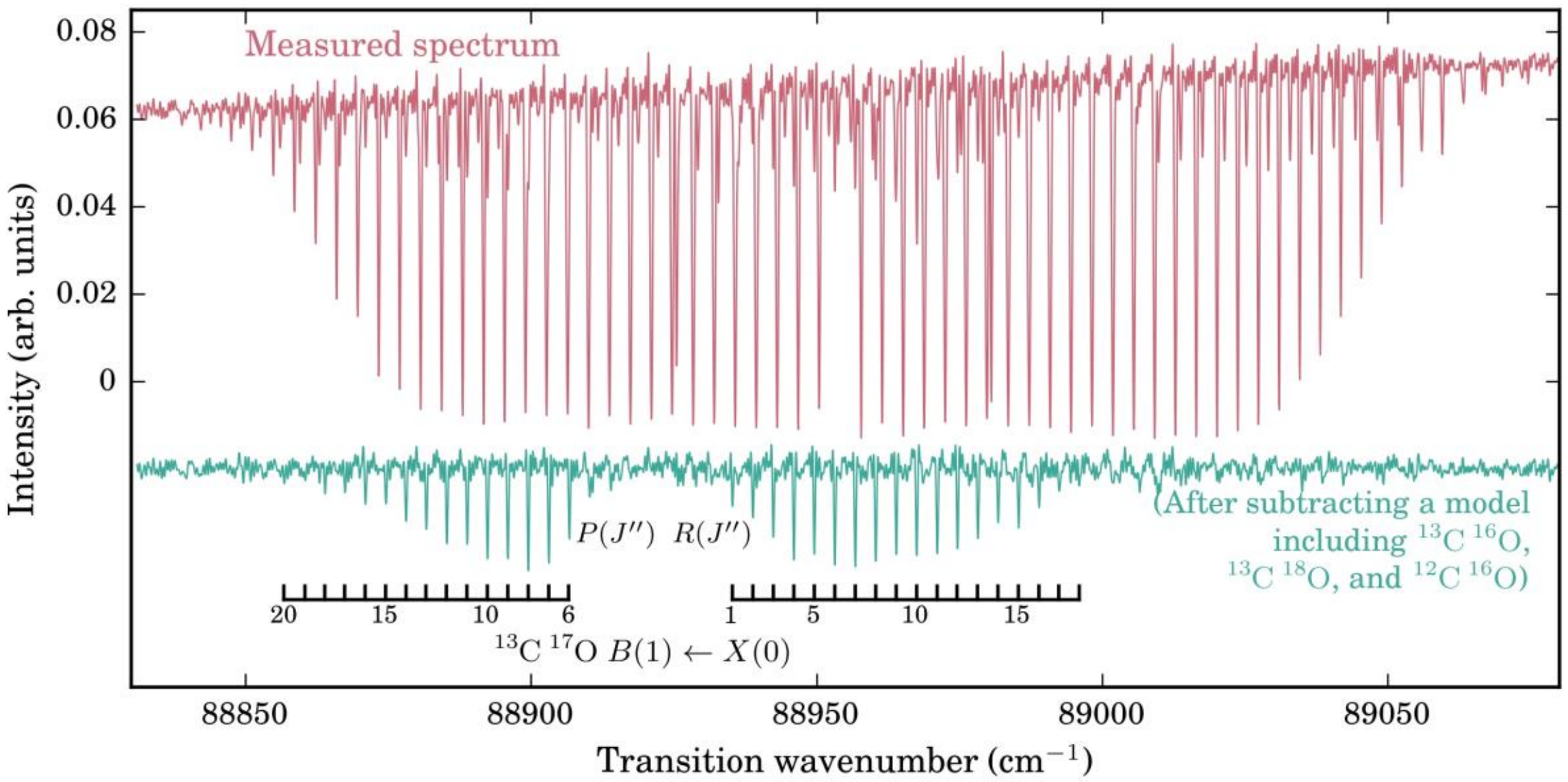

**Fig. 4.** Same caption as for Fig. 3 for the ($B^1\Sigma^+ \rightarrow X^1\Sigma^+$) (1, 0) band.

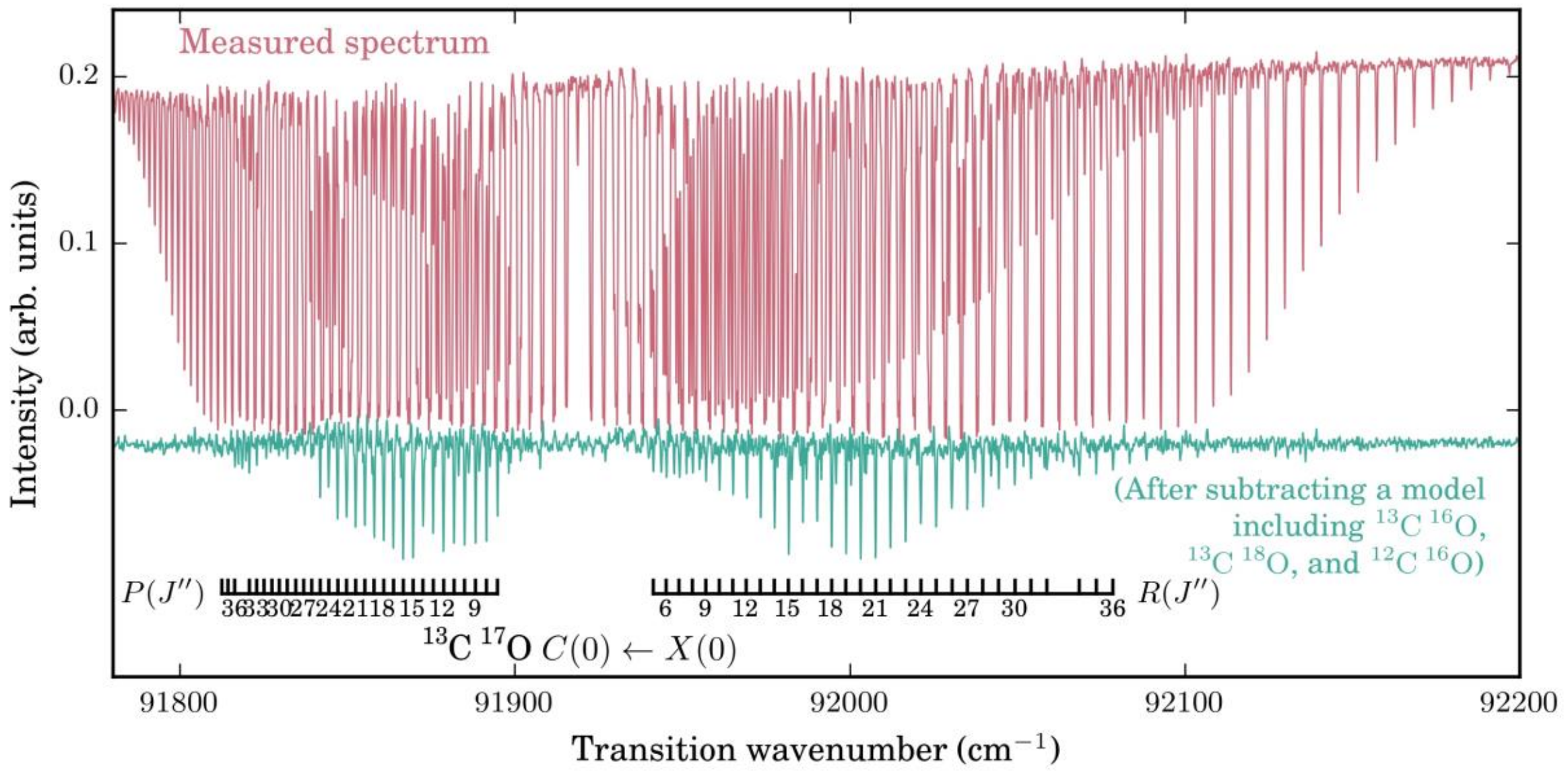

**Fig. 5.** Same caption as for Fig. 3 for the ($C^1\Sigma^+ \rightarrow X^1\Sigma^+$) (0, 0) band.

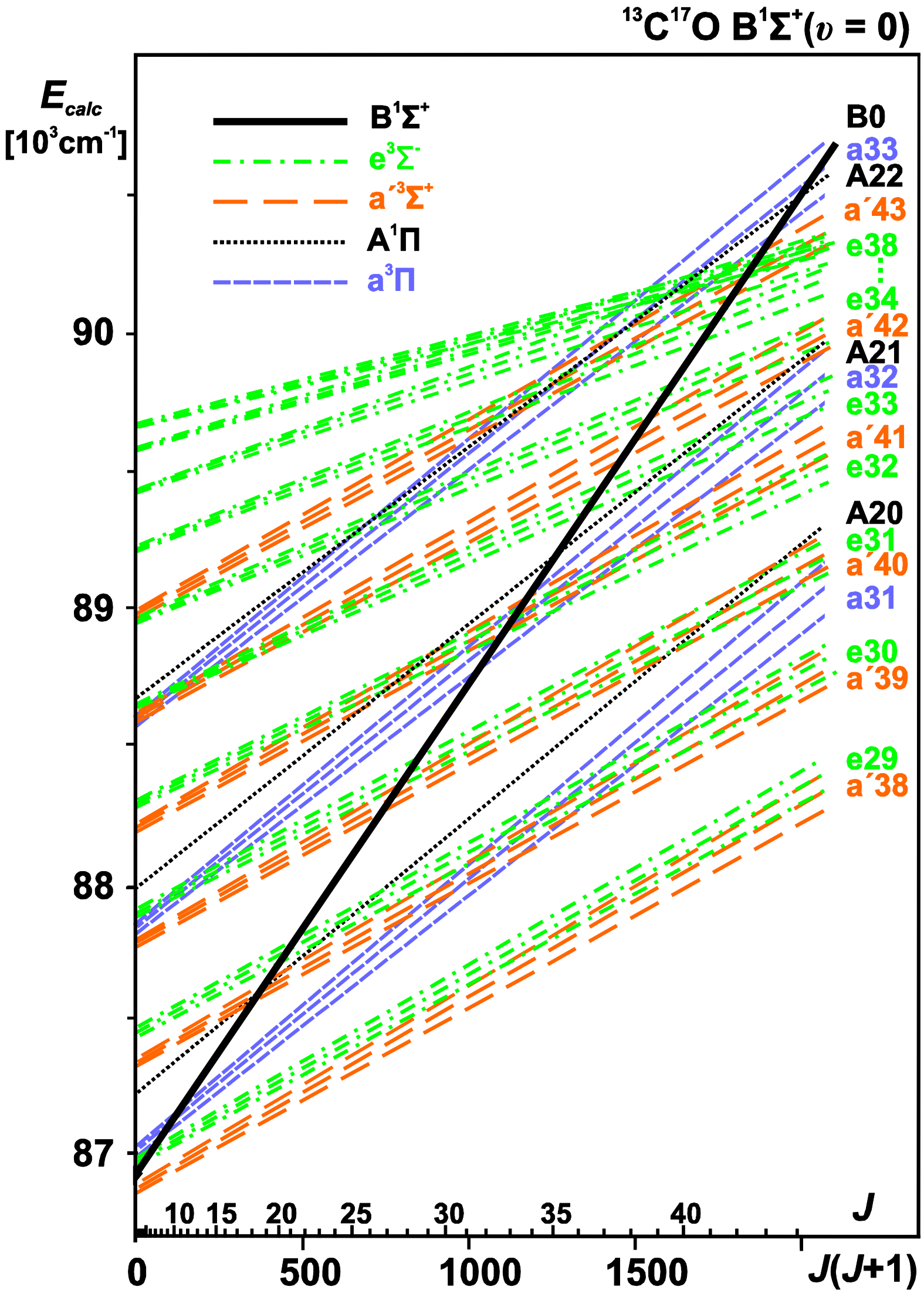

**Fig. 6.** Rovibronic term crossing diagram for the $^{13}C^{17}O$ $B^1\Sigma^+$ ($\upsilon$ = 0) level. The curves are indicated by an electronic state label and vibrational quantum number to their right.

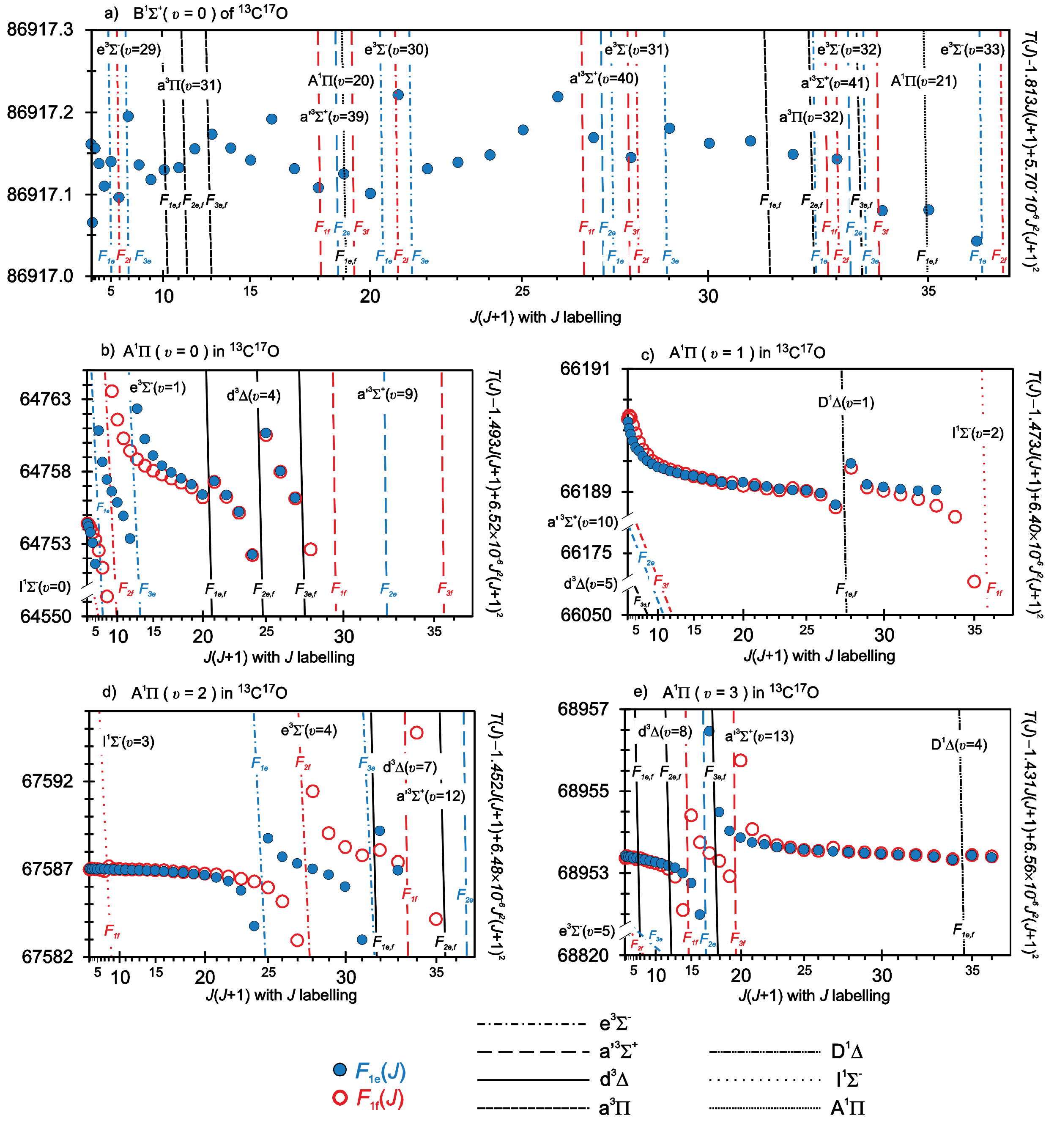

**Fig. 7.** The reduced term values (in cm$^{-1}$) plotted for the $^{13}C^{17}O$ $B^1\Sigma^+$ ($\upsilon$ = 0) and $A^1\Pi$ ($\upsilon$ = 0, 1, 2, 3) levels together with the hypothetical unperturbed crossing rovibronic levels of the possible perturbers. Filled (blue) and open (red) circles indicate *e* and *f* electronic symmetry of the B and A states, if applicable. In order to expand the vertical scale, $T(J) - BJ(J+1) + DJ^2(J+1)^2$ is subtracted from the observed term values (zero energy is placed at $\upsilon$ = 0, $J$ = 0 of the $X^1\Sigma^+$ ground state). Note that different reduced-energy scales are used for different vibrational levels.

**References**

[1] Bayet E, Gerin M, Phillips TG, Contursi A. A survey of submillimeter C and CO lines in nearby galaxies. Astron Astrophys 2006;460:467–85. doi:10.1051/0004-6361:20053872.
[2] Greve TR, Bertoldi F, Smail I, Neri R, Chapman SC, Blain AW, et al. An interferometric CO survey of luminous submillimetre galaxies. Mon Not R Astron Soc 2005;359:1165–83. doi:10.1111/j.1365-2966.2005.08979.x.
[3] Sham TK, Yang BX, Kirz J, Tse JS. K-edge near-edge X-ray-absorption fine structure of oxygen- and carbon-containing molecules in the gas phase. Phys Rev A 1989;40:652–69. doi:10.1103/PhysRevA.40.652.
[4] Salumbides EJ, Niu ML, Bagdonaite J, de Oliveira N, Joyeux D, Nahon L, et al. CO A−X system for constraining cosmological drift of the proton-electron mass ratio. Phys Rev A 2012;86:22510. doi:10.1103/PhysRevA.86.022510.
[5] De Nijs AJ, Salumbides EJ, Eikema KSE, Ubachs W, Bethlem HL. UV-frequency metrology on CO ($a^3\Pi$): Isotope effects and sensitivity to a variation of the proton-to-electron mass ratio. Phys Rev A 2011;84:52509. doi:10.1103/PhysRevA.84.052509.
[6] Tauber J, Ade PAR, Aghanim N, Alves MIR, Armitage-Caplan C, Arnaud M, et al. Planck 2013 results. XIII. Galactic CO emission. Astron Astrophys 2014;571. doi:10.1051/0004-6361/201321553.
[7] Sheffer Y, Rogers M, Federman SR, Abel NP, Gredel R, Lambert DL, et al. Ultraviolet survey of CO and $H_2$ in diffuse molecular clouds: The reflection of two photochemistry regimes in abundance relationships. Astrophys J 2008;687:1075–106. doi:10.1086/591484.
[8] Perez S, Casassus S, Ménard F, Roman P, Van Der Plas G, Cieza L, et al. CO gas inside the protoplanetary disk cavity in HD 142527: Disk structure from Alma. Astrophys J 2015;798. doi:10.1088/0004-637X/798/2/85.
[9] Sahai R, Mack-Crane GP. The astrosphere of the asymptotic giant branch star CIT 6. Astron J 2014;148. doi:10.1088/0004-6256/148/4/74.
[10] Catling DC. Planetary atmospheres. In: Shubert D (editor), Treatise on geophysics. vol. 10.13. 2nd ed. Oxford, Elsevier; 2015.
[11] Heng K, Lyons JR. Carbon dioxide in exoplanetary atmospheres: rarely dominant compared to carbon monoxide and water in hot, hydrogen-dominated atmospheres. Astrophys J 2016;817. doi:10.3847/0004-637X/817/2/149.
[12] Daprà M, Niu ML, Salumbides EJ, Murphy MT, Ubachs W. Constraint on a cosmological variation in the proton-to-electron mass ratio from electronic CO absorption. Astrophys J 2016;826:192. doi:10.3847/0004-637X/826/2/192.
[13] Stohl A, Berg T, Burkhart JF, Fjæraa AM, Forster C, Herber A, et al. Arctic smoke - record high air pollution levels in the European Arctic due to agricultural fires in Eastern Europe in spring 2006. Atmospheric Chem Phys 2007;7:511–34. doi:10.5194/acp-7-511-2007.
[14] Sumpf B, Burrows JP, Kissel A, Kronfeldt H-D, Kurtz O, Meusel I, et al. Line shift investigations for different isotopomers of carbon monoxide. J Mol Spectrosc 1998;190:226–31. doi:10.1006/jmsp.1998.7595.
[15] Ubachs W, Eikema KSE, Hogervorst W, Cacciani PC. Narrow-band tunable extreme-ultraviolet laser source for lifetime measurements and precision spectroscopy. J Opt Soc Am B 1997;14:2469–76. doi:10.1364/JOSAB.14.002469.
[16] Balat-Pichelin M, Iacono J, Boubert P. Recombination coefficient of atomic oxygen on ceramic materials in a $CO_2$ plasma flow for the simulation of a Martian entry. Ceram Int 2016;42:2761–9. doi:10.1016/j.ceramint.2015.11.007.
[17] Bally J, Stark AA, Wilson RW, Langer WD. Filamentary structure in the Orion molecular cloud. Astrophys J Lett 1987;312:L45–9. doi:10.1086/184817.
[18] Neininger N, Guelin M, Ungerechts H, Lucas R, Wielebinski R. Carbon monoxide emission as a precise tracer of molecular gas in the Andromeda galaxy. Nature 1998;395:871–873. doi:10.1038/27612.

[19] Schwarz KR, Bergin EA, Cleeves LI, Blake GA, Zhang K, Öberg KI, et al. The radial distribution of $H_2$ and CO in TW Hya as revealed by resolved ALMA observations of CO isotopologues. Astrophys J 2016;823. doi:10.3847/0004-637X/823/2/91.
[20] Visser R, van Dishoeck EF, Black JH. The photodissociation and chemistry of CO isotopologues: applications to interstellar clouds and circumstellar disks. Astron Astrophys 2009;503:323–43. doi:10.1051/0004-6361/200912129.
[21] Wannier PG, Lucas R, Linke RA, Encrenaz PJ, Penzias AA, Wilson RW. The abundance ratio [$^{17}$O]/[$^{18}$O] in dense interstellar clouds. Astrophys J 1976;205:L169. doi:10.1086/182116.
[22] Ladd EF, Fuller GA, Deane JR. C$^{18}$O and C$^{17}$O observations of embedded young stars in the Taurus Molecular Cloud. I. Integrated intensities and column densities. Astrophys J 1998;495:871–90. doi:10.1086/305313.
[23] Bensch F, Pak I, Wouterloot JGA, Klapper G, Winnewisser G. Detection of $^{13}$C$^{17}$O and observations of rare CO isotopomers toward the ρ Ophiuchi molecular cloud. Astrophys J Lett 2001;562:L185–8.
[24] Fuller GA, Ladd EF. The evolution of the circumstellar environment of embedded young stars from observations of rare species of carbon monoxide. Astrophys J 2002;573:699. doi:10.1086/340753.
[25] Sheffer Y, Lambert DL, Federman SR. Ultraviolet detection of interstellar $^{12}$C$^{17}$O and the CO isotopomeric ratios toward X Persei. Astrophys J 2002;574:L171–4. doi:10.1086/342501.
[26] Wouterloot JGA, Brand J, Henkel C. The interstellar C$^{18}$O / C$^{17}$O ratio in the solar neighbourhood: The ro-Ophiuchus cloud. Astron Astrophys 2005;430:549–60. doi:10.1051/0004-6361:20040437.
[27] Ladd EF. On the relative abundance of C$^{18}$O and C$^{17}$O in the Taurus Molecular Cloud. Astrophys J 2004;610:320–328. doi:10.1086/421387.
[28] Winnewisser G, Dumesh BS, Pak I, Surin LA, Lewen F, Roth DA, et al. Novel intracavity jet millimeter wave spectrometer: detection of b-type rotational transitions of Ne-CO. J Mol Spectrosc 1998;192:243–6. doi:10.1006/jmsp.1998.7672.
[29] Guelachvili G. Absolute wavenumbers and molecular constants of the fundamental bands of $^{12}$C$^{16}$O, $^{12}$C$^{17}$O, $^{12}$C$^{18}$O, $^{13}$C$^{16}$O, $^{13}$C$^{17}$O, $^{13}$C$^{18}$O and of the 2-1 bands of $^{12}$C$^{16}$O and $^{13}$C$^{16}$O, around 5 μm, by Fourier spectroscopy under vacuum. J Mol Spectrosc 1979;75:251–69. doi:10.1016/0022-2852(79)90121-8.
[30] Ubachs W, Velchev I, Cacciani P. Predissociation in the $E^1\Pi$, v=1 state of the six natural isotopomers of CO. J Chem Phys 2000;113:547–60. doi:10.1063/1.481830.
[31] Surin LA, Dumesh BS, Lewen F, Roth DA, Kostromin VP, Rusin FS, et al. Millimeter-wave intracavity-jet OROTRON-spectrometer for investigation of van der Waals complexes. Rev Sci Instrum 2001;72:2535–42. doi:10.1063/1.1369640.
[32] Lemaire JL, Eidelsberg M, Heays AN, Gavilan L, Federman SR, Stark G, et al. High-resolution spectroscopy of the $A^1\Pi$ (v'=0-10) - $X^1\Sigma^+$ (v''=0) bands in $^{13}$C$^{18}$O : term values, ro-vibrational oscillator strengths and Hönl–London corrections. J Phys B At Mol Opt Phys 2016;49:154001. doi:10.1088/0953-4075/49/15/154001.
[33] Hakalla R, Zachwieja M. Rotational analysis of the Ångström system ($B^1\Sigma^+ – A^1\Pi$) in the rare $^{13}$C$^{17}$O isotopologue. J Mol Spectrosc 2012;272:11–8. doi:10.1016/j.jms.2011.12.002.
[34] Hakalla R, Zachwieja M, Szajna W. First analysis of the 1– v″ progression of the Ångström ($B^1\Sigma^+ – A^1\Pi$) band system in the rare $^{13}$C$^{17}$O isotopologue. J Phys Chem A 2013;117:12299–312. doi:10.1021/jp4077239.
[35] Hakalla R. First analysis of the Herzberg ($C^1\Sigma^+ – A^1\Pi$) band system in the less-abundant $^{13}$C$^{17}$O isotopologue. RSC Adv 2014;4:44394–407. doi:10.1039/c4ra08222b.
[36] Lefebvre-Brion H, Field RW. The Spectra and Dynamics of Diatomic Molecules. Amsterdam, The Netherlands: Elsevier Academic Press; 2004.
[37] Krupenie PH. The band spectrum of carbon monoxide. vol. 5. National Standard

Reference Data Series, National Bureau of Standards, Washington, DC; 1966.
[38] Simmons JD, Bass AM, Tilford SG. The fourth positive system of carbon monoxide observed in absorption at high resolution in the vacuum ultraviolet region. Astrophys J 1969;155:345–58. doi:10.1086/149869.
[39] Kępa R, Rytel M. The Ångström ($B^1\Sigma^+$ - $A^1\Pi$) system of the CO molecules: New observations and analyses. J Phys B 1993;26:3355–62. doi:10.1088/0953-4075/26/19/023.
[40] Kępa R. High-resolution studies of the Herzberg band system ($C^1\Sigma^+$ – $A^1\Pi$) in the $^{13}C^{18}O$ molecule. Can J Phys 1988;66:1012–24. doi:10.1139/p88-163.
[41] Ostrowska-Kopeć M, Piotrowska I, Kępa R, Kowalczyk P, Zachwieja M, Hakalla R. New observations and analyses of highly excited bands of the fourth-positive ($A^1\Pi \rightarrow X^1\Sigma^+$) band system in $^{12}C^{16}O$. J Mol Spectrosc 2015;314:63–72. doi:10.1016/j.jms.2015.06.004.
[42] Field RW. Ph.D. Thesis. Harvard University, 1971.
[43] Field RW, Wicke BG, Simmons JD, Tilford SG. Analysis of perturbations in the $a^3\Pi$ and $A^1\Pi$ states of CO. J Mol Spectrosc 1972;44:383–99. doi:10.1016/0022-2852(72)90111-7.
[44] Le Floch AC, Launay F, Rostas J, Field RW, Brown CM, Yoshino K. Reinvestigation of the CO $A^1\Pi$ state and its perturbations: the v = 0 level. J Mol Spectrosc 1987;121:337–79. doi:10.1016/0022-2852(87)90056-7.
[45] Le Floch AC. Ph.D. Thesis. University Paris-Sud, 1989.
[46] Le Floch A. Accurate energy levels for the $C^1\Sigma^+$ (v = 0) and $E^1\Pi$ (v = 0) states of $^{12}C^{16}O$. J Mol Spectrosc 1992;155:177–83. doi:10.1016/0022-2852(92)90557-5.
[47] Niu ML, Salumbides EJ, Zhao D, de Oliveira N, Joyeux D, Nahon L, et al. High resolution spectroscopy and perturbation analysis of the CO $A^1\Pi - X^1\Sigma^+$ (0,0) and (1,0) bands. Mol Phys 2013;111:2163–74. doi:10.1080/00268976.2013.793889.
[48] Niu ML, Salumbides EJ, Heays AN, de Oliveira N, Field RW, Ubachs W. Spectroscopy and perturbation analysis of the CO $A^1\Pi - X^1\Sigma^+$ (2,0), (3,0) and (4,0) bands. Mol Phys 2016;114:627–36. doi:10.1080/00268976.2015.1108472.
[49] Haridass C, Huber KP. A high-resolution $^{13}C$ isotope study in the vacuum ultraviolet of spectra of CO(A→X), C I, and C II. Astrophys J 1994;420:433–8. doi:10.1086/173573.
[50] Gavilan L, Lemaire JL, Eidelsberg M, Federman SR, Stark G, Heays AN, et al. High-resolution study of $^{13}C^{16}O$ A–X($v'$=0–9) bands using the VUV-FTS at SOLEIL: revised term values. J Phys Chem A 2013;117:9644–52. doi:10.1021/jp312338z.
[51] Niu ML, Hakalla R, Madhu Trivikram T, Heays AN, De Oliveira N, Salumbides EJ, et al. Spectroscopy and perturbation analysis of the $A^1\Pi$ (v = 0) state of $^{13}C^{16}O$. Mol Phys 2016;accepted for publication. doi:10.1080/00268976.2016.1218078.
[52] Haridass C, Reddy SP, Le Floch AC. The fourth positive ($A^1\Pi$ - $X^1\Sigma^+$) system of $^{12}C^{18}O$ and $^{13}C^{18}O$: perturbations in the $A^1\Pi$ state. J Mol Spectrosc 1994;167:334–52. doi:10.1006/jmsp.1994.1240.
[53] Haridass C, Reddy SP, Le Floch AC. Precise Rovibronic Term Values of Some Vibrational Levels of the $A^1\Pi$, $B^1\Sigma^+$, $C^1\Sigma^+$, and $E^1\Pi$ States of $^{12}C^{18}O$ and $^{13}C^{18}O$. J Mol Spectrosc 1994;168:429–41. doi:10.1006/jmsp.1994.1291.
[54] Beaty LM, Braun VD, Huber KP, Le Floch AC. A high-resolution $^{18}O$ isotope study in the vacuum ultraviolet of the $A^1\Pi \rightarrow X^1\Sigma^+$ 4$^{th}$ positive system of CO. Astrophys J Suppl Ser 1997;109:269–77. doi:10.1086/312976.
[55] Hakalla R, Niu M, Field RW, Salumbides E, Heays A, Stark G, et al. VIS and VUV spectroscopy of $^{12}C^{17}O$ and deperturbation analysis of the $A^1\Pi$, υ = 1 – 5 levels. RSC Adv 2016;6:31588–606. doi:10.1039/C6RA01358A.
[56] Amiot C, Roncin J-Y, Verges J. First observation of the CO $E^1\Pi$ - $B^1\Sigma^+$ and $C^1\Sigma^+$ - $B^1\Sigma^+$ band systems. Predissociation in the $E^1\Pi$ (v=0) level. J Phys B 1986;19:L19. doi:10.1088/0022-3700/19/1/004.
[57] Janjić JD, Danielak J, Kępa R, Rytel M. The Ångström bands of $^{13}C^{16}O$ and $^{12}C^{18}O$ molecules. Acta Phys Pol A 1972;41:757–61.
[58] Lemaire JL, Eidelsberg M, Heays AN, Gavilan L, Stark G, Federman SR, et al. New

and revised high-resolution spectroscopy of six CO isotopologues in the 101-115 nm range (Transition energies, Term values and Molecular constants of the B, C, and E states). Forthcom Publ 2016.
[59] Bacis R. A new source for the search and study of electronic molecular spectra: the composite wall hollow cathode (CWHC). J Phys E Sci Instrum 1976;9:1081–6. doi:10.1088/0022-3735/9/12/021.
[60] Levenberg K. A method for the solution of certain non-linear problems in least squares. Q Appl Math 1944;2:164–8.
[61] Marquardt DW. An algorithm for least-squares estimation of nonlinear parameters. J Soc Ind Appl Math 1963;11:431–41.
[62] OPUS spectroscopy software, v. 7.5.18. Bruker Optik GmbH; 2014.
[63] Kępa R, Rytel M. On the Ångström bands of $^{12}C^{16}O$. Acta Phys Pol A 1970;37:585–90.
[64] Le Floch AC, Amiot C. Fourier transform spectroscopy of the CO Ångström bands. Chem Phys 1985;97:379–89.
[65] de Oliveira N, Roudjane M, Joyeux D, Phalippou D, Rodier J-C, Nahon L. High-resolution broad-bandwidth Fourier-transform absorption spectroscopy in the VUV range down to 40 nm. Nat Photonics 2011;5:149–53. doi:10.1038/nphoton.2010.314.
[66] de Oliveira N, Joyeux D, Phalippou D, Rodier JC, Polack F, Vervloet M, et al. A Fourier transform spectrometer without a beam splitter for the vacuum ultraviolet range: From the optical design to the first UV spectrum. Rev Sci Instrum 2009;80:43101. doi:10.1063/1.3111452.
[67] Nahon L, De Oliveira N, Garcia GA, Gil J-F, Pilette B, Marcouillé O, et al. DESIRS: a state-of-the-art VUV beamline featuring high resolution and variable polarization for spectroscopy and dichroism at SOLEIL. J Synchrotron Radiat 2012;19:508–20. doi:10.1107/S0909049512010588.
[68] de Oliveira N, Joyeux D, Roudjane M, Gil J-F, Pilette B, Archer L, et al. The high-resolution absorption spectroscopy branch on the VUV beamline DESIRS at SOLEIL. J Synchrotron Radiat 2016;23:887–900. doi:10.1107/S1600577516006135.
[69] Drabbels M, Heinze J, ter Meulen JJ, Meerts WL. High resolution double-resonance spectroscopy on Rydberg states of CO. J Chem Phys 1993;99:5701. doi:10.1063/1.465919.
[70] Cacciani P, Brandi F, Velchev I, Lyngå C, Wahlström C-G, Ubachs W. Isotope dependent predissociation in the $C^1\Sigma^+$, v= 0 and v= 1 states of CO. Eur Phys J D 2001;15:47–56. doi:10.1007/s100530170182.
[71] Kovács I. Rotational structure in the spectra of diatomic molecules. London, England: Adam Hilger Ltd.; 1969.
[72] Kronig R de L. Zur deutung der bandenspektren II. Z Phys 1928;50:347–62. doi:10.1007/BF01347513.
[73] Havenith M, Bohle W, Werner J, Urban W. Vibration rotation spectroscopy of excited electronic states Faraday-L.M.R. spectroscopy of CO $a^3\Pi$. Mol Phys 1988;64:1073–88. doi:10.1080/00268978800100723.
[74] Kittrell C, Garetz BA. Analysis of the $D^1\Delta$ - $X^1\Sigma^+$ transition in CO observed by two-photon excitation. Spectrochim Acta 1989;45:31–40. doi:10.1016/0584-8539(89)80024-8.
[75] Tilford SG, Simmons JD. Atlas of the observed absorption spectrum of carbon monoxide between 1060 and 1900 Å. J Phys Chem Ref Data 1972;1:147–88. doi:10.1063/1.3253097.
[76] Huber KP, Herzberg G. Constants of diatomic molecules. New York: Van Nostrand Reinhold; 1979.
[77] Coxon JA, Hajigeorgiou PG. Direct potential fit analysis of the $X^1\Sigma^+$ ground state of CO. J Chem Phys 2004;121:2992–3008. doi:10.1063/1.1768167.
[78] Kępa R, Ostrowska-Kopeć M, Piotrowska I, Zachwieja M, Hakalla R, Szajna W, et al. Ångström ($B^1\Sigma^+ \rightarrow A^1\Pi$) 0-1 and 1-1 bands in isotopic CO molecules: further investigations. J Phys B 2014;47:45101. doi:10.1088/0953-4075/47/4/045101.

[79] Western CM. PGOPHER, a program for simulating rotational, vibrational and electronic structure of spectra. University of Bristol: 2016.
[80] Bergeman T, Cossart D. The lower excited states of CS: a study of extensive spin-orbit perturbations. J Mol Spectrosc 1981;87:119–95. doi:10.1016/0022-2852(81)90088-6.
[81] Dunham JL. The energy levels of a rotating vibrator. Phys Rev 1932;41:721–31. doi:10.1103/PhysRev.41.721.
[82] Hall J.A., Schamps J., Robbe J.M., Lefebvre-Brion H. Theoretical study of the perturbation parameters in the $a^3\Pi$ and $A^1\Pi$ states of CO. J Chem Phys 1973;59:3271–83. doi:10.1063/1.1680469.
[83] Jung C. FRACON-Programmdokumentation Quantenchemie. Berlin: Akademie der Wissenschaften der DDR; 1979.
[84] Jakubek Z. FRACONB-Programmdokumentation Quantenchemie FRACON (ver. B). Rzeszów: Pedagogical University of Rzeszów; 1988.
[85] Curl RF, Dane CB. Unbiased least-squares fitting of lower states. J Mol Spectrosc 1988;128:406–12. doi:10.1016/0022-2852(88)90157-9.
[86] Watson JKG. On the use of term values in the least-squares fitting of spectra. J Mol Spectrosc 1989;138:302–8. doi:10.1016/0022-2852(89)90119-7.
[87] Heays AN, Niu ML, Lemaire JL, Eidelsberg M, Federman SR, Stark G, et al. High-resolution photoabsorption spectra of $B^1\Sigma^+$ - $X^1\Sigma^+(0; 0)$ recorded at the SOLEIL synchrotron. 2017:in preparation.
[88] Morton DC, Noreau L. A compilation of electronic transitions in the CO molecule and the interpretation of some puzzling interstellar absorption features. Astrophys J Suppl Ser 1994;95:301–44. doi:10.1086/192100.
[89] Le Floch A. Revised molecular constants for the ground state of CO. Mol Phys 1991;72:133–44. doi:10.1080/00268979100100081.
[90] Field RW, Tilford SG, Howard RA, Simmons JD. Fine structure and perturbation analysis of the $a^3\Pi$ state of CO. J Mol Spectrosc 1972;44:347–82. doi:10.1016/0022-2852(72)90110-5.
[91] Herzberg G, Simmons JD, Bass AM, Tilford SG. Forbidden $I^1\Sigma^-$ - $X^1\Sigma^+$ absorption bands of carbon monoxide. Can J Phys 1966;44:3039–45. doi:10.1139/p66-249.

# Supplementary material

## 1. Effective Hamiltonian and matrix elements

Effective Hamiltonian and matrix elements used in deperturbation analyses of the $A^1\Pi$ ($v$ = 0 - 3) levels in $^{13}C^{17}O$ [a, b, c, d, e]

| | $A^1\Pi$ | $I^1\Sigma^-$ | $D^1\Delta$ | $e\,^3\Sigma^-$ | $a'\,^3\Sigma^+$ | $d\,^3\Delta$ |
|---|---|---|---|---|---|---|
| $A^1\Pi$ | $T_v + \left(B \pm \frac{q}{2}\right)\hat{\boldsymbol{N}}^2 - D\hat{\boldsymbol{N}}^4 + H\hat{\boldsymbol{N}}^6$ | $\xi_i(I_v) \times \left(\hat{\boldsymbol{N}}_+\hat{\boldsymbol{L}}_- + \hat{\boldsymbol{N}}_-\hat{\boldsymbol{L}}_+\right)$ | $\xi_i(D_v) \times \left(\hat{\boldsymbol{N}}_+\hat{\boldsymbol{L}}_- + \hat{\boldsymbol{N}}_-\hat{\boldsymbol{L}}_+\right)$ | $\eta_i(e_v)\hat{\boldsymbol{L}}\cdot\hat{\boldsymbol{S}}$ | $\eta_i(a'_v)\hat{\boldsymbol{L}}\cdot\hat{\boldsymbol{S}}$ | $\eta_i(d_v)\hat{\boldsymbol{L}}\cdot\hat{\boldsymbol{S}}$ |
| $I^1\Sigma^-$ | | $T_v + B\hat{\boldsymbol{N}}^2 - D\hat{\boldsymbol{N}}^4 + H\hat{\boldsymbol{N}}^6$ | 0 | 0 | 0 | 0 |
| $D^1\Delta$ | | | $T_v + B\hat{\boldsymbol{N}}^2 - D\hat{\boldsymbol{N}}^4 + H\hat{\boldsymbol{N}}^6$ | 0 | 0 | 0 |
| $e\,^3\Sigma^-$ | | | | $T_v + B\hat{\boldsymbol{N}}^2 - D\hat{\boldsymbol{N}}^4 + H\hat{\boldsymbol{N}}^6 + \frac{2}{3}\lambda\left(3\hat{\boldsymbol{S}}_z^2 - \hat{\boldsymbol{S}}^2\right)$ | 0 | 0 |
| $a'\,^3\Sigma^+$ | | | | | $T_v + B\hat{\boldsymbol{N}}^2 - D\hat{\boldsymbol{N}}^4 + H\hat{\boldsymbol{N}}^6 + \frac{2}{3}\lambda\left(3\hat{\boldsymbol{S}}_z^2 - \hat{\boldsymbol{S}}^2\right) + \gamma\left(\hat{\boldsymbol{N}}\cdot\hat{\boldsymbol{S}}\right)$ | 0 |
| $d\,^3\Delta$ | | | | | | $T_v + B\hat{\boldsymbol{N}}^2 - D\hat{\boldsymbol{N}}^4 + H\hat{\boldsymbol{N}}^6 + \frac{2}{3}\lambda\left(3\hat{\boldsymbol{S}}_z^2 - \hat{\boldsymbol{S}}^2\right) + \gamma\left(\hat{\boldsymbol{N}}\cdot\hat{\boldsymbol{S}}\right) + A\hat{\boldsymbol{L}}_z\hat{\boldsymbol{S}}_z + \frac{1}{2}A_D\lambda\left(\hat{\boldsymbol{N}}^2\hat{\boldsymbol{L}}_z\hat{\boldsymbol{S}}_z + \hat{\boldsymbol{L}}_z\hat{\boldsymbol{S}}_z\hat{\boldsymbol{N}}^2\right)$ |

[a] Matrix elements of the Hamiltonian are similar to Field [RW. Ph.D. Thesis. Harvard University, 1971], Bergeman et al. [J Mol Spectrosc 1981;87:119–95], and Le Floch et al. [J Mol Spectrosc 1987;121:337–79], retaining their symbols for the various molecular constants.

[b] The matrix is symmetric, thus the left-sided, non-diagonal elements, not shown in the Hamiltonian, are equivalent to the right-sided, non-diagonal ones.

[c] The mutual interactions between the perturbing states are neglected (the matrix elements set to zero).

[d] The '+' and '–' signs for the $A^1\Pi$ diagonal element indicate *e*- and *f*- symmetry levels, respectively.

[e] $T_v$ – means rotation-less energies in relation to the $X^1\Sigma^+(v = 0)$ level; $\eta_i$ and $\xi_i$ – indicate spin-orbit and *L*-uncoupling interaction parameter, respectively.

## 2. Perturbation parameters

If $\boldsymbol{H}^{\mathbf{SO}}$ and $\boldsymbol{H}^{\mathbf{RE}}$ are the spin-orbit and rotation-electronic operators, respectively, the effective perturbation parameters $\alpha$ and $\beta$, in the *e*/*f* basis set, are defined as follows [Le Floch et al. J Mol Spectrosc 1987;121:337–79]:

$$\alpha_{\mathrm{A\sim d}} = \langle \mathrm{A}^1\Pi,\, \upsilon_{\mathrm{A}} \left|\mathbf{H}^{\mathbf{SO}}\right| \mathrm{d}^3\Delta, \upsilon_{\mathrm{d}} \rangle = -\left(\tfrac{\sqrt{2}}{4}\right)\mathbf{a}_{\mathbf{A\sim d}}\langle \upsilon_{\mathrm{A}}|\upsilon_{\mathrm{d}}\rangle, \quad (1)$$

$$\alpha_{\mathrm{A\sim e}} = \langle \mathrm{A}^1\Pi,\, \upsilon_{\mathrm{A}} \left|\mathbf{H}^{\mathbf{SO}}\right| \mathrm{e}^3\Sigma^-, \upsilon_{\mathrm{e}} \rangle = -\left(\tfrac{1}{4}\right)\mathbf{a}_{\mathbf{A\sim e}}\langle \upsilon_{\mathrm{A}}|\upsilon_{\mathrm{e}}\rangle, \quad (2)$$

$$\alpha_{\mathrm{A\sim a'}} = \langle \mathrm{A}^1\Pi,\, \upsilon_{\mathrm{A}} \left|\mathbf{H}^{\mathbf{SO}}\right| \mathrm{a'}^3\Sigma^+, \upsilon_{\mathrm{a'}} \rangle = \left(\tfrac{1}{4}\right)\mathbf{a}_{\mathbf{A\sim a'}}\langle \upsilon_{\mathrm{A}}|\upsilon_{\mathrm{a'}}\rangle, \quad (3)$$

$$2\beta_{\mathrm{A\sim I}}\sqrt{\mathrm{x}} = \langle \mathrm{A}^1\Pi,\, \upsilon_{\mathrm{A}} \left|\mathbf{H}^{\mathbf{RE}}\right| \mathrm{I}^1\Sigma^-, \upsilon_{\mathrm{I}} \rangle = -\sqrt{\mathrm{x}}\,\mathbf{b}_{\mathbf{A\sim I}}\langle \upsilon_{\mathrm{A}}|\mathbf{B}(\mathbf{R})|\upsilon_{\mathrm{I}}\rangle, \quad (4)$$

$$\beta_{\mathrm{A\sim D}}\sqrt{\mathrm{x}-2} = \langle \mathrm{A}^1\Pi,\, \upsilon_{\mathrm{A}} \left|\mathbf{H}^{\mathbf{RE}}\right| \mathrm{D}^1\Delta, \upsilon_{\mathrm{D}} \rangle = \sqrt{\mathrm{x}-2}\,\mathbf{b}_{\mathbf{A\sim D}}\langle \upsilon_{\mathrm{A}}|\mathbf{B}(\mathbf{R})|\upsilon_{\mathrm{D}}\rangle, \quad (5)$$

where $x = \sqrt{J(J+1)}$, $\boldsymbol{a}_{\mathrm{A\sim d,e,a'}}$ and $\boldsymbol{b}_{\mathrm{A\sim I,D}}$ denote isotopologue-independent spin-orbit and rotation-electronic perturbation parameters, $\langle \upsilon_A | \upsilon_{d,e,a'} \rangle$ - vibrational overlap integrals, $\langle \upsilon_A | B(R) | \upsilon_{I,D} \rangle$ - rotational operator integrals.

The relationships between the perturbation parameters $\alpha_i$, $\beta_{\mathrm{A\sim I}}$, $\beta_{\mathrm{A\sim D}}$ (eqns. 1-5) and $\eta_i$, $\xi_{\mathrm{A\sim I}}$, $\xi_{\mathrm{A\sim D}}$ (used in this work) result from their different definitions:

$$\eta_i = -\alpha_i\sqrt{3}, \quad (6)$$

$$\xi_{A\sim I} = -\beta_{A\sim I}\sqrt{2}, \quad (7)$$

$$\xi_{A\sim D} = \beta_{A\sim D}, \quad (8)$$

where subscript '$i$' indicates A ~ d, A ~ e, as well as A ~ a' interactions.